\documentclass[11pt]{article}
\usepackage{times}
\usepackage{sidecap}
\usepackage{geometry}
\usepackage{floatrow}
\usepackage{float}
\usepackage{hyperref}
\usepackage{graphics,graphicx}
\usepackage{subcaption}
\usepackage{epsf,wrapfig}
\usepackage{microtype}

\usepackage{cleveref}

\usepackage{natbib,bm}

\usepackage{paralist}
\let\olditem\item
\renewenvironment{thebibliography}[1]{%
  \section*{\refname}
  \let\par\relax
  \renewcommand{\item}[1][]{\olditem[\textbullet]}%
  \inparaenum}{\endinparaenum}

\newcounter{aff}

\DeclareGraphicsExtensions{.jpg,.pdf,.png,.eps}

\usepackage{caption}
\usepackage{xcolor}

\usepackage{tcolorbox}

\bibpunct{(}{)}{;}{a}{}{,}

\usepackage{soul}

\definecolor{darkcyan}{rgb}{0.0, 0.0, 0.7}

\definecolor{boxcolor}{RGB}{193, 229, 255}
\definecolor{figboxcolor}{RGB}{193, 229, 255}
\definecolor{boxcolorgray}{RGB}{227, 227, 227}
\definecolor{darkcyan}{rgb}{0.0, 0.0, 0.7}

\sethlcolor{boxcolorgray}

\def\gtrsim{\lower 2pt \hbox{$\, \buildrel {\scriptstyle >}\over
{\scriptstyle \sim}\,$}}
\def\lesssim{\lower 2pt \hbox{$\, \buildrel {\scriptstyle <}\over
{\scriptstyle \sim}\,$}}

\def\approxlt{\lower.2em\hbox{$\buildrel < \over \sim$}}
\def\approxgt{\lower.2em\hbox{$\buildrel > \over \sim$}}

\def\be{\begin{equation}}
\def\ee{\end{equation}}
\def\bea{\begin{eqnarray*}}
\def\eea{\end{eqnarray*}}

\def\jref@jnl#1{{\rm#1}}
\def\aj{\jref@jnl{AJ}}			
\def\araa{\jref@jnl{ARA\&A}}		
\def\apj{\jref@jnl{ApJ}}			
\def\apjl{\jref@jnl{ApJ}}		
\def\apjs{\jref@jnl{ApJS}}		
\def\ao{\jref@jnl{Appl.~Opt.}}		
\def\apss{\jref@jnl{Ap\&SS}}		
\def\aap{\jref@jnl{A\&A}}		
\def\aapr{\jref@jnl{A\&A~Rev.}}		
\def\aaps{\jref@jnl{A\&AS}}		
\def\azh{\jref@jnl{AZh}}			
\def\baas{\jref@jnl{BAAS}}		
\def\jrasc{\jref@jnl{JRASC}}		
\def\memras{\jref@jnl{MmRAS}}		
\def\mnras{\jref@jnl{MNRAS}}		
\def\pra{\jref@jnl{Phys.~Rev.~A}}	
\def\prb{\jref@jnl{Phys.~Rev.~B}}	
\def\prc{\jref@jnl{Phys.~Rev.~C}}	
\def\prd{\jref@jnl{Phys.~Rev.~D}}	
\def\pre{\jref@jnl{Phys.~Rev.~E}}	
\def\prl{\jref@jnl{Phys.~Rev.~Lett.}}	
\def\pasp{\jref@jnl{PASP}}		
\def\pasj{\jref@jnl{PASJ}}		
\def\qjras{\jref@jnl{QJRAS}}		
\def\skytel{\jref@jnl{S\&T}}		
\def\solphys{\jref@jnl{Sol.~Phys.}}	
\def\sovast{\jref@jnl{Soviet~Ast.}}	
\def\ssr{\jref@jnl{Space~Sci.~Rev.}}	
\def\zap{\jref@jnl{ZAp}}			
\def\nat{\jref@jnl{Nature}}		
\def\iaucirc{\jref@jnl{IAU~Circ.}}
\def\aplett{\jref@jnl{Astrophys.~Lett.}}
\def\apspr{\jref@jnl{Astrophys.~Space~Phys.~Res.}}
\def\bain{\jref@jnl{Bull.~Astron.~Inst.~Netherlands}}
\def\fcp{\jref@jnl{Fund.~Cosmic~Phys.}}
\def\gca{\jref@jnl{Geochim.~Cosmochim.~Acta}}
\def\grl{\jref@jnl{Geophys.~Res.~Lett.}}
\def\jcp{\jref@jnl{J.~Chem.~Phys.}}	
\def\jgr{\jref@jnl{J.~Geophys.~Res.}}	
\def\jqsrt{\jref@jnl{J.~Quant.~Spec.~Radiat.~Transf.}}
\def\memsai{\jref@jnl{Mem.~Soc.~Astron.~Italiana}}
\def\nphysa{\jref@jnl{Nucl.~Phys.~A}}
\def\physrep{\jref@jnl{Phys.~Rep.}}
\def\physscr{\jref@jnl{Phys.~Scr}}
\def\planss{\jref@jnl{Planet.~Space~Sci.}}
\def\procspie{\jref@jnl{Proc.~SPIE}}

\def\ion#1#2{#1$\;${\small\rm\@Roman{#2}}\relax}

\newcount\lecurrentfam
\def\LaTeX{\lecurrentfam=\the\fam \leavevmode L\raise.42ex
\hbox{$\fam\lecurrentfam\scriptstyle\kern-.3em A$}\kern-.15em\TeX}
\def\sizrpt{
(\fontname\the\font): em=\the\fontdimen6\font, ex=\the\fontdimen5\font
\typeout{
(\fontname\the\font): em=\the\fontdimen6\font, ex=\the\fontdimen5\font
}}
\definecolor{dark}{rgb}{ 0.34, 0.09, 0.49}

\usepackage{fancyhdr,url}
\usepackage{transparent}
\usepackage{tablefootnote}
\usepackage{mathpazo}

\begin{document}

\noindent
\begin{raggedright}
{\Huge\bfseries The transformative power of ALMA2040 for mm-VLBI  \\[0.5em]
\Large From black-hole snapshots to population studies, precision tests of gravity, cosmology, masers, and the time-domain Universe
}\end{raggedright}\\

\noindent
The Event Horizon Telescope Collaboration
\vspace{1em}
{
\footnotesize{
\noindent 
}}

\begin{tcolorbox}[boxsep=3pt,left=0pt,right=0pt,top=1pt,bottom=1pt,colframe=boxcolor,colback=boxcolor]
\justify
\textbf{The field of millimeter very long baseline interferometry (mm-VLBI) is continuously growing and has a clear roadmap for future instrumentation upgrades. The current focus of mm-VLBI arrays is on multi-frequency observations. With the Event Horizon Telescope (EHT), we want to employ multi-frequency measurements to resolve fainter emission regions and perform precision tests of gravity after our initial images of the ring-like emission in M87* and Sgr~A*. Here, our VLBI array will benefit greatly from the ALMA2040 project. Phasing up the full ALMA2040 array will provide simultaneous ``multi-color'' views of the event-horizon region with a full collecting area of an equivalent 225-m mm dish, allowing us to disentangle effects of gravity, magnetic fields, and optical depth on horizon-scales as we can utilize the full simultaneous frequency coverage between ALMA bands 2 and 9 with the future EHT. The future Global mm-VLBI Array (GMVA) will extend the mm-VLBI frequency range from 3\,mm down to 17mm, benefiting primarily from the ALMA bands 1\,\&\,2.\\
In this work, we quantify the expected differences in science return between the capabilities of ALMA with its Wideband Sensitivity Upgrade (WSU) versus ALMA2040 for mm-VLBI in the areas of gravitational and black hole physics, cosmology, masers, jet physics, microquasars, as well as rapid transient follow-up. 
We find that ALMA2040 provides us with sub-mJy baseline sensitivities at 230\,GHz and enables 345\,GHz imaging of faint sources, providing us with the prospect of moving horizon-scale science from two individual black holes towards population studies of black hole shadows across mass, accretion state, and environments. We also expect a further improvement in studies with the intra-ALMA data enabled by the VLBI observations: Inferences of polarized hotspots in accretion flows, searches for pulsars in the Galactic Center and deep transient searches in general. Combined with rapid-response capabilities across the VLBI network, ALMA2040 will help transform the EHT from an instrument producing individual snapshots into a true time-domain observatory.\\
With ALMA2040, we can get a high-fidelity multi-frequency reconstruction of the jet in M87* down to the ring-like horizon-scale emission. This provides a way to investigate the connection between the innermost accretion flow and the launching region of the relativistic jet.
For Sgr~A*, we get a meaningful improvement when deconvolving the interstellar scattering screen, which leads to a 27\,\% reduction in the error of the recovered intrinsic source structure,  bringing the unscattered structure of Sgr A* at lower frequencies within reach. .} 
\end{tcolorbox}

\newpage{}

\vspace{-0.7cm}
\section{Introduction}

We have assembled the Event Horizon Telescope (EHT) as a global very long baseline interferometry (VLBI) array to primarily image black holes on horizon scales. Thus far, we have imaged the shadows cast by the supermassive black holes M87* \citep{M87p1,M87p2,M87p3,M87p4,M87p5,M87p6,M87p8,M87p9, M87_2018p1} and Sgr~A* \citep{SgrA_2022p1,SgrA_2022p2,SgrA_2022p3,SgrA_2022p4,SgrA_2022p5,SgrA_2022p6, SgrA_2024p7,SgrA_2024p8} as direct evidence for the existence of supermassive black holes and tests of General Relativity in the unique regime of strong gravitational potential and small spacetime curvature.

Through continued observations from a constantly upgraded and expanded array, we narrow down the spectral and polarimetric properties of the horizon-scale and extended emissions of our primary targets and measure their temporal evolution. Together with multi-wavelength observations \citep{2021MWL}, we constrain the plasma and accretion dynamics, enabling stringent tests of fundamental physics.
Observing at wavelengths of 1.3 and 0.87\,mm with a maximum resolving power of 15\,$\mu$as, we also image the launching region of extragalactic radio jets \citep{2020Kim, 2021Janssen, 2023Jorstad, 2024Baczko, 2024Paraschos, 2025Roeder, 2026Mueller}. Besides tests of gravity, we probe dynamic accretion and plasma processes at the highest achievable resolving power, jointly with coordinated multi-wavelength campaigns \citep{2021MWL} to constrain emission processes and connect our images to the large-scale structure of the observed sources.
Phased ALMA plays a crucial role for the EHT as the most sensitive antenna in the array, significantly boosting our fringe detection thresholds \citep{2019MsngrGoddi}. The VLBI mode makes efficient use of ALMA time as shown by \citet{2026Stoehr}.

While we focus on phased ALMA2040 joining the EHT in this article, we also note that the enormous collecting area and simultaneous multi-frequency capabilities between the 43 and 86\,GHz VLBI bands (ALMA bands 1 and 2) will greatly benefit future observations of the GMVA \citep{2024Ros}, whose planned Frequency Phase Transfer (FPT)-enabling upgrade to a 22/43/86\,GHz global VLBI array \citep{2023Dodson} will allow phase transfer from the lower frequencies to 86\,GHz for an improved sensitivity and astrometry of compact structures, and the ngVLA \citep{2018Selina} with further VLBI extensions like the Long-baseline Extension in next-generation VLBI Experiments and
Rapid-response Array Germany/Europe (LEVERAGE) project \citep{2026Neidhardt}.
Finally, we note the great benefit for space mm-VLBI missions that connect to a ground array, for which multiple concepts are currently being pursued by various groups. The expected scientific return of such space missions can be significantly expanded to fainter sources \citep[e.g., being able to detect many more black hole shadows and photon ring signatures around black holes with the extreme resolving power, e.g.,][]{2021Pesce} with an ALMA2040 array.

We have organized our white paper as follows: \Cref{sec:technical} describes the technical requirements for VLBI, including studies for the expected improvements for future ALMA2040+EHT observations. The most important aspect will be the ability to phase up the full ALMA2040 array simultaneously at multiple frequency bands.
\Cref{sec:science} provides a bird's eye view on the expected scientific studies enabled by ALMA2040+EHT, including multi-messenger synergies within time-domain astrophysics.

\section{ALMA2040 Instrumentation}
\label{sec:technical}

\subsection{Technical requirements for VLBI}

The ALMA2040 project allows for technical upgrades that will yield significant improvements for VLBI observations. These are
\begin{enumerate}
    \item An upgraded phasing system \citep{2018Matthews} with a fast phasing loop $\mathcal{O}$(s) to enable phasing up all ALMA dishes, out to longer baselines (current ALMA limits are set at $\sim$ 500m). A few larger individual dishes, as suggested in some of the ALMA2040 design concepts, could help as reference antennas for the phasing on weak sources. With the current phasing system, we would only be able to phase up a central compact core of antennas that only have short baselines, where we might also suffer from shadowing.
    \item Longer intra-ALMA baselines of up to 32\,km to narrow the gap between the shortest EHT baselines of 160\,m and 2.6\,km and the next shortest baselines of 100\,km, enabling the reconstruction of source structures at currently inaccessible scales, combined by feathering. 
    \item The ability to phase up all ALMA bands at the same time for frequency phase transfer~\citep[FPT, see, e.g.,][]{Rioja2011,Issaoun2025,Zhao2025}, Faraday tomography, source spectral evolution and opacity studies, broadband spectroscopy on $\mu$as scales, and removal of the interstellar scattering screen \citep[e.g.,][]{Johnson2018} towards Sgr~A*. For a few select ALMA2040 dishes, that could serve as phasing system reference antennas, installing pulse cal systems would greatly improve the effectiveness and stability of FPT \citep{2022Janssen}.
    \item New dishes with good aperture efficiencies at 690\,GHz. While VLBI remains challenging at those frequencies \citep{2026Chen}, obtaining long-baseline detections for model-fitting and the imaging of bright sources would become feasible with ALMA2040.
    
    \item The ability to enter VLBI mode within minutes at any time for rapid target of opportunity observations, aimed at transients.
    \item The ability to form multiple phasing centers within the primary beam of the ALMA dishes for astrometric studies, notably for the Galactic Center region. This will require sufficient recording media for the different summed signals.
    \item A frequency setup for the recorded VLBI data that matches those of the other mm-VLBI antennas. This eases the correlation without the loss of sub-channels through zoom-bands \citep{2022Janssen}.
\end{enumerate}

\subsection{Sensitivity}

We consider a simple observation on a long baseline to phased ALMA and a dual-frequency setup between 86\,GHz (3\,mm) and 230\,GHz (1.3\,mm). 
The 3\,mm band is used to solve phases on a short atmospheric coherence timescale. Phase solutions are scaled and applied to the 1.3\,mm band via FPT to extend the coherence at the short wavelength over long integration times~\citep{Rioja2023,Issaoun2025,Zhao2025}.
For simplicity, we assume the correlated (resolved) flux densities $S_\nu$ at different observing frequencies on the same baseline to be related only by a spectral index $\alpha$: $S_\nu \propto \nu^{-\alpha}$. With $\sigma$ the thermal noise, we can then write the expected signal-to-noise ratio $\xi$ as
\begin{equation}
    \xi = \frac{S_\nu}{\sigma} = S_\nu \eta_q \eta_c \sqrt{\frac{2 \Delta \nu \Delta \mathrm{t}}{\mathrm{SEFD}_1 \mathrm{SEFD}_2}} \; ,
\end{equation}
with $\eta_q =0.881$ being the quantization efficiency for 2-bit sampling, $\Delta \nu$ the recorded bandwidth, $\Delta \mathrm{t}$ the integration time, $\eta_c$ the coherence factor over $\Delta$t, and the system equivalent flux densities (SEFDs)  as the total noise budgets of the two telescopes connecting the baseline.

\begin{SCfigure}
    \centering
    \includegraphics[width=0.5\linewidth]{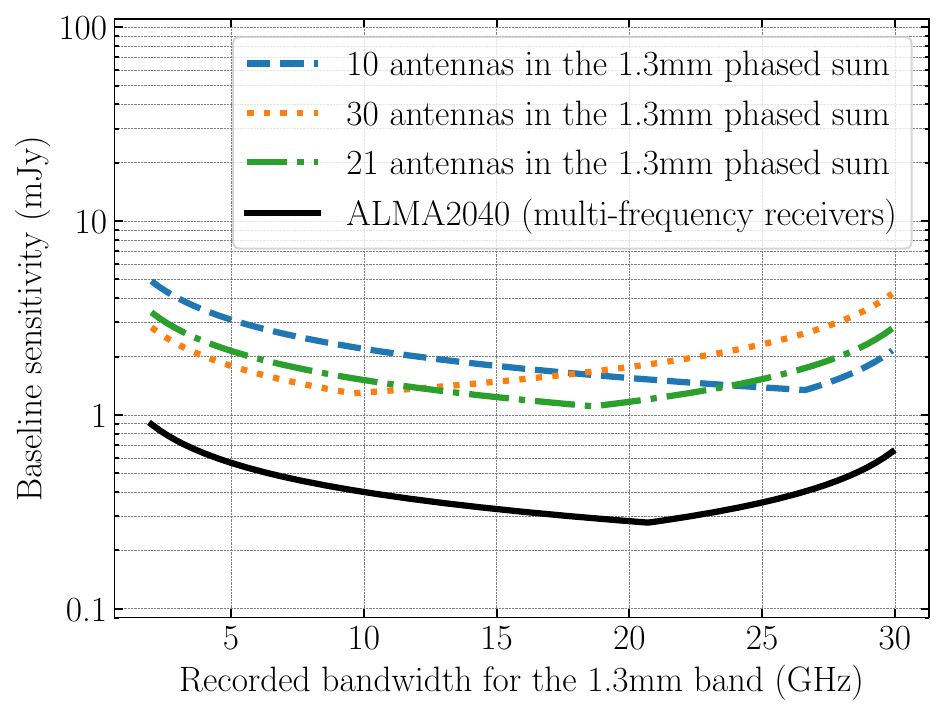}
    \caption{Sensitivity to the correlated (resolved) source flux density on long baselines (from the LMT or NOEMA) to ALMA with FPT, comparing WSU to ALMA2040. The available 32\,GHz bandwidth is split between the 1.3\,mm ``science'' and the 3\,mm ``FPT calibration'' bands. ALMA2040 can phase up fully at all bands. In the WSU configuration, separate sub-arrays are formed from 37 antennas, with an optimum sensitivity for 16 antennas phased up at 3\,mm and 21 at 1.3\,mm in this case study. Baselines to a phased ALMA2040 reach sub-mJy baseline sensitivity at 230 GHz, a 4× gain over ALMA WSU, opening faint sources to horizon-scale imaging.}
    \label{fig:sensitivity} 
\end{SCfigure}

For the phase-steering at 3\,mm, we require $\xi|_{3\,mm}$ = 3 over a typical coherence time of $\Delta \mathrm{t}|_{3\,mm}$ = 15\,s, and $\alpha=0.5$ as a conservative estimate.
For the 1.3\,mm band, we require a robust fringe detection with $\xi|_{1.3\,mm}$ = 7 over a full VLBI scan duration of $\Delta \mathrm{t}|_{1.3\,mm} = 300$\,s and we assume FPT to work with $\eta_c|_{1.3\,mm} = 0.9$. 
The SEFDs nominally depend on the local atmospheric conditions and source elevation. Here, we take typical values of 700\,Jy and 1000\,Jy for the sensitive stations that provide long baselines to ALMA (the LMT and NOEMA) at 3\,mm and 1.3\,mm, respectively. 
For phased ALMA, we scale the SEFDs according to the number of antennas in the phased sum, with baseline values for 37 dishes of 70\,Jy and 100\,Jy at 3\,mm and 1.3\,mm, respectively.
Following this case study, \Cref{fig:sensitivity} depicts an \textit{approximately fourfold sensitivity gain for the assumed observing configuration with ALMA2040 compared to ALMA with the Wideband Sensitivity Upgrade (WSU).}\footnote{A similar improvement can be expected for the future GMVA with FPT for 3\,mm.}

\begin{figure}
    \centering
    \includegraphics[width=\linewidth]{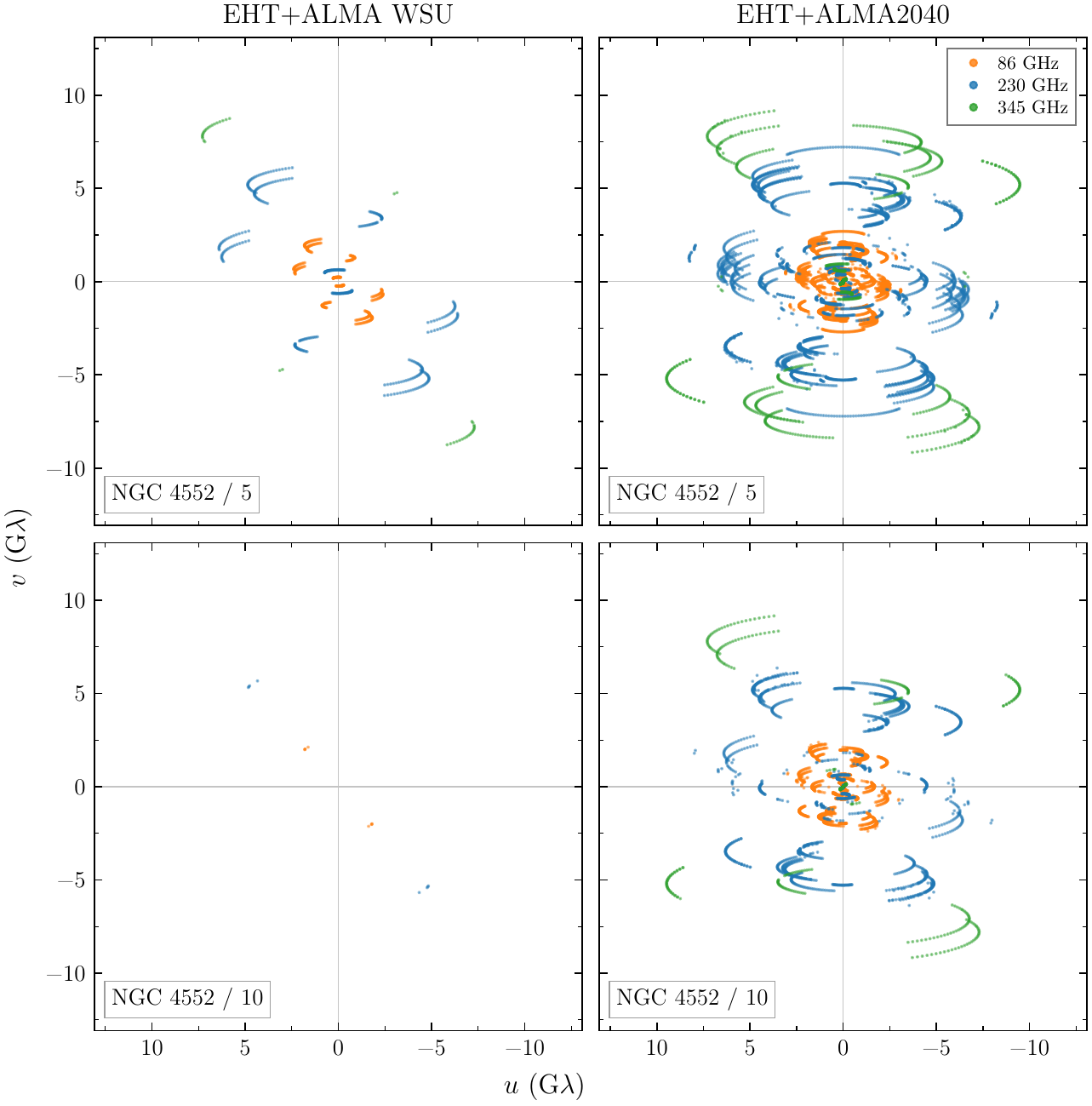}
    \caption{EHT baseline coverages with S/N$>$5 detections from simulated observations of a weak source for EHT plus ALMA WSU on the left and ALMA2040 on the right. We took the NGC~4552 model from \citet{2025Zhang}, which has total flux densities of $\sim$21, $\sim$36, and $\sim$37 mJy at 86, 230, and 345 GHz, respectively, and lowered those flux densities by a factor of 5 in the top panel and a factor of 10 in the bottom panel. Here, joint observations with the future GMVA would further extend the 86\,GHz coverage and expand the coverage also to 22 and 43\,GHz. Through the common 3\,mm band, we would have an astrometric alignment from 22 to 345\,GHz. These simulations illustrate how ALMA2040 could extend detectable VLBI emission to sources 5--10$\times$ fainter than the adopted NGC~4552 model, opening new opportunities for multi-frequency imaging and population studies.} 
    \label{fig:NGC4552}
\end{figure}

In \Cref{fig:NGC4552} we present simulated observations following \citet{2024Pesce}, which show significant improvements in the expected $(u,v)$-coverages due to ALMA2040 when observing faint sources. Thus, we will be able to image a larger population of sources \citep{2023Ramakrishnan,2024Nair}, potentially turning black-hole shadow detection from a handful of individual case studies into population science in addition to several other science advances enumerated in \Cref{sec:science} below.

\section{Scientific Context and Open Questions}
\label{sec:science}

In this section, we briefly discuss the various future science cases for the EHT and how those would benefit from ALMA2040. For more comprehensive overviews, we refer to \citet{2024midrangescience} for our short-term goals and \citet{2023Johnson, 2025Ayzenberg} for a description of future goals enabled by instrumental upgrades to our array. In the 2040s time frame, we expect the EHT to be able to record 32\,GHz of aggregate bandwidth and to be enhanced by the Africa Millimetre Telescope \citep{2016Backes}, the Tenerife EHT Antenna (Gomez et al. 2026, in press) as well as several additional ngEHT telescopes \citep{2023Doeleman}. In addition, there are  current investigations for mm-wave telescopes in Germany, Switzerland, Tanzania, South Africa, and the Himalayan region.\newpage

Simulated images in this work are produced via general-relativistic radiative transfer (GRRT) post-processing of general-relativistic magneto-hydrodynamics (GRMHD) simulations \citep{Younsi2020,Chatterjee2025}.
Our model for a faint future EHT target (\Cref{fig:NGC4552}) is described in \citet{2025Zhang}.
Our model for M87* (\Cref{fig:M87_2040}) is described in \citet{2025Tsunetoe}. 

\subsection{Strong-field gravity}

{\bf Expanding the horizon-scale black-hole sample:} By resolving the horizon-scale emission around supermassive black holes, the EHT probes gravity in the hitherto unexplored regime of strong gravitational potential and relatively weak spacetime curvature \citep{SgrA_2022p6}.
With the ALMA2040 sensitivity enabling higher angular resolution observations at 345\,GHz and 690\,GHz as central VLBI reference station, we can expand our sample of detectable black hole shadows from two sources to weaker, more distant, and lower-mass systems \citep{2021Pesce}. Tapping into a larger sample will allow us to distinguish horizon-scale properties from accretion state and environmental effects.

{\bf Precision inference of black hole astrophysics:} Through multi-frequency imaging \citep[e.g.,][]{2017Moscibrodzka, 2023Ricarte} and by sheer reduction of thermal noise on baselines to ALMA \citep{2025Janssen}, we expect significant improvements in comparisons between our observations and the next-generation of GRMHD simulations and alternative gravity models.
We will be able to simultaneously constrain spacetime metrics, the spin of the central compact object, plasma physics, and mm-wave emission mechanisms.

{\bf Precision strong-field gravity:}  
ALMA2040+EHT will improve tests of strong-field gravity by providing higher-fidelity measurements of horizon-scale structure, particularly at the highest observing frequencies where the emission is less affected by optical-depth and scattering effects. More precise measurements of the size and shape of the lensed emission can tighten constraints on deviations from the Kerr geometry beyond those currently obtained from EHT observations of Sgr~A* \citep{SgrA_2022p6}. Time-resolved imaging will additionally allow us to test the stability of these structures and separate persistent geometric features from variability in the accretion flow. Polarimetric imaging provides complementary information: the morphology and evolution of the polarized emission can constrain the magnetic-field structure and accretion dynamics \citep{SgrA_2024p8}, while also carrying information about black-hole spin \citep{2023Ricarte_spin} and, potentially, higher-order multipolar structures. Together, these observables provide several 
complementary ways of testing the Kerr spacetime rather than relying on the shadow size alone.

\subsection{AGN physics}

{\bf Relativistic plasma physics and magnetic fields} \citep[see][for a review]{2019Blandford}{\bf :} Simultaneous multi-frequency VLBI observations at 86/230/345\,GHz will provide unprecedented measurements of Faraday rotation, circular polarization, and depolarization, allowing us to determine the magnetic-field structure, plasma magnetization, and particle composition (electron--ion versus pair plasma) close to the event horizon. 

\begin{figure}[h]
    \centering
    \includegraphics[width=\linewidth]{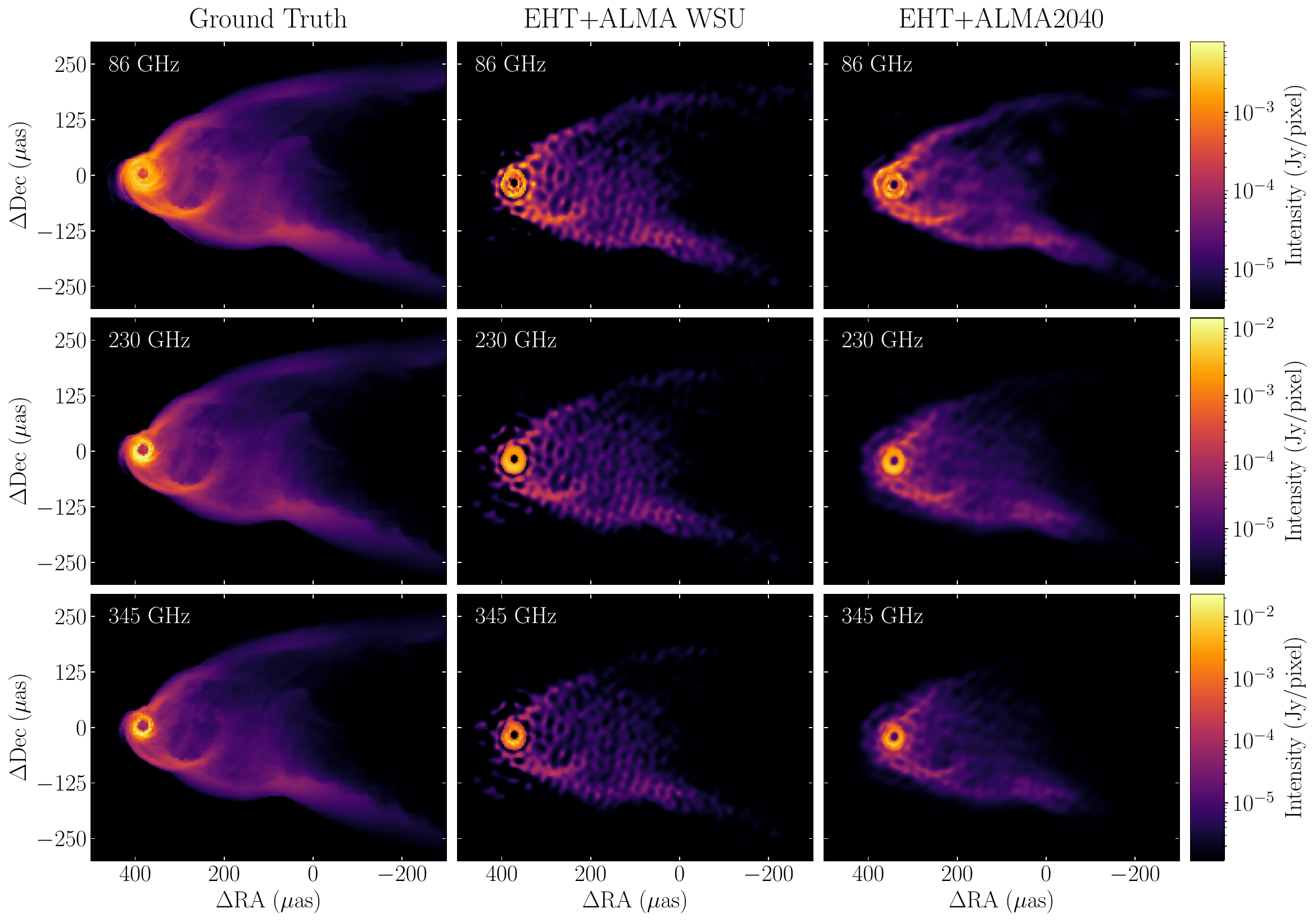}
    \caption{Multi-frequency reconstructions of synthetic data sets 
    with \texttt{eht-imaging} \citep{2023Chael}. The M87 image models are taken from \citet{2025Tsunetoe}. The reconstructions compare the impact of ALMA2040 compared to ALMA WSU joining future EHT observations.
    ALMA2040 provides an excellent super-resolution to connect also the very small-scale jet structures to the horizon-scale properties of the accretion and completely eliminates all imaging artifacts.
    }
    \label{fig:M87_2040}
\end{figure}

{\bf Jet-launching physics} \citep[see][for a review]{2017Boccardi}{\bf :} At larger scales, for extragalactic radio jets launched by AGN, ALMA2040+EHT will enable detailed studies of jet formation, acceleration, collimation, composition, core shifts, edge-brightening, couplings to accretion flows, and magnetic-field structures, as well as strengths over a broad range of black-hole masses and accretion states, thanks to the ALMA2040 sensitivity boost. With mm-VLBI, our AGN observations will zoom in on the Poynting-flux-dominated jet launching region. Secular changes will be studied through multi-epoch monitoring over years to decades, to shed light on the life cycle of supermassive black holes. The future 22/43/86 GHz VLBI observations will then allow us to connect to the jet morphologies and emission properties at larger scales.
In \Cref{fig:M87_2040}, we show the improvements in our ability to image the base of the M87 jet, connecting to the ring-like horizon-scale emission and thus linking the accretion flow properties directly to the jet launching. With ALMA2040, we can get a \textit{high-fidelity reconstruction of the jet and we can clearly identify the small-scale structures within the jet at all frequencies, thanks to a strong super-resolution.}

{\bf Scattering studies and screen recoveries in the Galactic Center.} The strong scattering that affects the GC limits the study and image reconstruction of sources such as Sgr A*, in particular for frequencies below 230\,GHz~\citep{Bower2006,Johnson2018,Issaoun2019,Issaoun2021, Cho2022}. A high-sensitivity station in a large VLBI array would improve the mitigation results and even allow us to reconstruct the scattering stochastic screen situated at approximately halfway between the Galactic center and the observer~\citep{Johnson2016, Mus2025}.
To demonstrate the capabilities of ALMA2040 in addressing the challenging problem of interstellar scattering, we have simulated 86\,GHz EHT observations of Sgr~A* with the appropriate scattering kernel. Applying a multi-objective optimization algorithm \citep{Mus2025} to these data, we find that the \textit{RMS of the intrinsic source structure recovery improves by 27\,\%} when ALMA2040 is in the array, compared to ALMA WSU. This brings the intrinsic, unscattered structure of Sgr A* at lower frequencies within reach, where it is otherwise blurred by the interstellar screen.

\subsection{Cosmology}

{\bf Improving \(H_0\) measurements from AGN disk masers:} VLBI monitoring of masers in AGN disks can be used to model their orbits around the central black hole, which permits astronomers to derive geometric distances to these AGN out to over 100 Mpc \citep[e.g.,][]{Kuo2013,Kuo2015,Gao2016}. Across multiple sources, these distance measurements can be used to constrain the Hubble constant in a manner that is independent of standard candles or distance ladders \citep{Pesce2020}. (Sub)millimeter maser transitions offer an opportunity to achieve an order-of-magnitude improvement in angular resolution and correspondingly more precise VLBI maps -- and thus distance measurements -- for these systems. There are several hundred known AGN megamaser systems emitting at 22\,GHz \citep{Kuo2020}, though to date only a couple dozen have been observed to emit in (sub)millimeter transitions \citep[e.g.,][]{Pesce2023}.  Presumably many more such systems exist at a detectable level, with the current population statistics limited primarily by a lack of systematic surveys. ALMA2040 can increase both the number of measurable maser components within individual disks and the number of systems accessible to such measurements. This will provide stronger constraints on disk models and their geometric distances, reducing both statistical uncertainties and source-specific systematics in an independent determination of \(H_0\). 

\subsection{Stellar maser science}

{\bf Formation of high-mass stars:} The mm/submm regime contains several maser transitions arising in (slow and fast) shocks and irradiation regions in the surroundings of a forming star. VLBI spectropolarimetric mapping reaches AU scales, revealing the detailed structure, velocity- and magnetic field of the gas in the disk and in the wind and outflow \citep{Sanna2017,Moscadelli2022}. With ALMA2040, it will be possible to phase up multiple spectral windows across multiple bands allowing simultaneous, astrometrically registered maps of multiple maser transitions that can capture the infall and outflow of individual gas streams. Such combined mapping of this variable emission will have the power to correlate and constrain the mass growth and outflow kinematics in young stars.

{\bf Mass-loss in evolved stars:} Strong stellar winds from evolved stars lead to a chemical enrichment of the gas in a galaxy. With different maser species, we can investigate different layers of these winds, for example, SiO masers come from closer to the stellar surface than water masers \citep{Yoon2018N}. The Abundance of SiO masers in the inner Galaxy \citep[e.g.,][]{2018Trapp} allows for precision astrometry and dynamics studies of the Milky Way bar and bulge, where optical astronomy is hindered by strong extinction. VLBI of mm/submm maser transitions has the angular resolution (of 0.001~$R_\star$ for a distance of 200~pc) to resolve single maser clouds. Also here, the power of ALMA2040 lies in the simultaneous, astrometrically registered mapping of different maser species from different wind layers to reveal the structure and kinematics of the mass outflow at various $R_\star$. 

\subsection{Time-domain astrophysics at the highest spatial resolution}

We plan to have an agile EHT array in the future, to respond quickly to transient events and resolve variable structures on $\mu$as scales, where ALMA2040 allows us to target a larger population of faint sources.

{\bf Direct observations of the accretion flow and jet coupling} \citep{1982Blandford}{\bf :} The ALMA2040 sensitivity will help us detect faint outflow/jet emission beyond horizon scales.
Over timescales from hours for Sgr~A* and days for M87*, we can probe the dynamical relation between AGN inflows and outflows; in particular, the spatially and temporally resolved effects of plasma dynamics, turbulence, magnetic reconnection, and flare evolution.

{\bf Transients and multi-wavelength variability:} Following AGNs and stellar-mass black holes (XRBs) that display multi-wavelength flaring activity \citep[e.g.,][]{1997Ulrich, 2004Fender}, "changing-look" characteristics \citep{2023Ricci}, or are generally in an outburst phase, we can follow up with ALMA2040+EHT $\mu$as-resolution mm-wave observations to reveal the connections between particle acceleration, magnetic energy dissipation, plasma heating, and structural changes in the emitting region.
Additionally, we expect to be able to resolve the jets from other transient sources such as gamma-ray bursts, tidal disruption events, and supernovae \citep[e.g.,][]{2022Eftekhari}.
We thus have a natural synergy with observatories that study high-energy astrophysical events and time-domain facilities. While other facilities typically provide only light-curves, we will be able to spatially resolve the structure of the emission, which will yield unique model constraints.
Complementing the current generation of gravitational-wave detectors, we envision follow-up observations of neutron-star merger events \citep{2017Abbott}.
Studies of binary supermassive black-hole systems \citep[e.g.,][]{2024Zhao} can be coordinated with future space-based gravitational-wave detectors.
Coordination with neutrino observatories will allow us to resolve the jet launching region of candidate AGN counterparts \citep{2023Kovalev}.

{\bf Synergies with ALMA2040 connected-element data:} During VLBI observations, we also obtain the intra-ALMA interferometric data with long integration times needed for VLBI. These enable transient searches for orbiting hotspots \citep[Q--U loops,][]{2022Wielgus} that can be used to infer spin, searching for pulsars around Sgr~A* that would enable precision tests of GR \citep{2023Torne}, as well as general mm-wave transients \citep{2022Mus}. The sensitivity improvement of ALMA2040 will enable significantly deeper searches here.

\section{Summary}

The role of ALMA in mm/submm VLBI goes well beyond the sensitivity provided by a single station. ALMA2040 can make high-frequency VLBI substantially more robust and extend it to sources, frequencies, and timescales that are difficult to access today. In particular, improved sensitivity and phasing, simultaneous multi-band observations, and rapid response work together to enable the detection of weaker sources, transfer phase information between frequencies, follow structural and polarimetric changes across bands, and respond to transient or rapidly evolving events. 
These capabilities are especially important as horizon-scale science moves beyond individual images. Repeated and multi-frequency observations will make it possible to follow the evolution of the emitting plasma and magnetic field, connect the innermost accretion flow to the formation of relativistic jets, and improve tests of the spacetime around black holes. At the same time, the larger accessible source population will help determine which properties are common to black-hole systems and which depend on mass, accretion state, environment, or source geometry. The same improvements will strengthen a broader range of VLBI science, including maser astrometry and geometric distances, stellar environments, and studies of interstellar scattering. 

ALMA2040 will also become increasingly important as part of a broader observational ecosystem. Its multi-frequency capabilities naturally complement the GMVA and future global mm/submm VLBI arrays, while rapid-response observations provide a direct connection to time-domain and multi-messenger facilities. In the longer term, the sensitivity of phased ALMA will be particularly valuable for VLBI baselines to new ground-based stations and potential space-VLBI elements. In this context, ALMA2040 provides the sensitivity needed to turn the angular resolution of a global array into measurements of the physics of black holes, relativistic plasmas, jets, and other compact astrophysical systems.

\newpage
\section*{Acknowledgements}

This work was developed as part of the broader ALMA2040 community effort. We acknowledge the Lorentz Center and NOVA for supporting the community workshops and coordination activities that helped initiate and shape the ALMA2040 White Paper process.

\sloppy
The Event Horizon Telescope Collaboration thanks the following
organizations and programs: the Academia Sinica; the Research Council of Finland (project 362572); the Agencia Nacional de Investigaci\'{o}n y Desarrollo (ANID), Chile via NCN$19\_058$ (TITANs), Fondecyt 1221421 and 11251078, and BASAL FB210003; the Alexander
von Humboldt Stiftung (including the Feodor Lynen Fellowship); an Alfred P. Sloan Research Fellowship;
the ALMA North America Development Fund; the Astrophysics and High Energy Physics programme by MCIN (with funding from European Union NextGenerationEU, PRTR-C17I1); the Black Hole Initiative, which is funded by grants from the John Templeton Foundation (60477, 61497, 62286) and the Gordon and Betty Moore Foundation (Grants GBMF-8273, GBMF12987) -- although the opinions expressed in this work are those of the authors and do not necessarily reflect the views of these Foundations; 
the Brinson Foundation; the Canada Research Chairs (CRC) program; 
the University of Toronto Eric and Wendy Schmidt AI in Science Postdoctoral Fellowship Program, a program of Schmidt Sciences;
the Natural Sciences \& Engineering Research Council of Canada (NSERC);
the Ontario Research Fund - Research Excellence (Project Number RE012-045);
Chandra DD7-18089X and TM6-17006X; the China Scholarship
Council; the China Postdoctoral Science Foundation fellowships (2020M671266, 2022M712084); ANID through Fondecyt Postdoctorado (project 3250762); Conicyt through Fondecyt Postdoctorado (project 3220195); Consejo Nacional de Humanidades, Ciencia y Tecnología (CONAHCYT, Mexico, projects U0004-246083, U0004-259839, F0003-272050, M0037-279006, F0003-281692, 104497, 275201, 263356, CBF2023-2024-1102, 257435, Ph.D. Scholarship 963437); the Colfuturo Scholarship; the Consejo Superior de Investigaciones 
Cient\'{i}ficas (grant 2019AEP112);
the Delaney Family via the Delaney Family John A.
Wheeler Chair at Perimeter Institute; Dirección General de Asuntos del Personal Académico-Universidad Nacional Autónoma de México (DGAPA-UNAM, projects IN112820 and IN108324); the Dutch Research Council (NWO) for the VICI award (grant 639.043.513), the grant OCENW.KLEIN.113, and the Dutch Black Hole Consortium (with project No. NWA 1292.19.202) of the research programme the National Science Agenda; the Dutch National Supercomputers, Cartesius and Snellius  (NWO grant 2021.013); 
the EACOA Fellowship awarded by the East Asia Core
Observatories Association, which consists of the Academia Sinica Institute of Astronomy and Astrophysics, the National Astronomical Observatory of Japan, Center for Astronomical Mega-Science,
Chinese Academy of Sciences, and the Korea Astronomy and Space Science Institute; 
the European Research Council (ERC) Synergy Grant ``BlackHoleCam: Imaging the Event Horizon of Black Holes'' (grant 610058), Synergy Grant ``BlackHolistic:  Colour Movies of Black Holes:
Understanding Black Hole Astrophysics from the Event Horizon to Galactic Scales'' (grant 10107164), and Horizon ERC Grants 2021 program under grant agreement No. 101040021; 
the European Union Horizon 2020
research and innovation programme under grant agreements
BlackHolistic (No. 101071643), 
RadioNet (No. 730562), 
M2FINDERS (No. 101018682); the European Research Council for advanced grant ``JETSET: Launching, propagation and 
emission of relativistic jets from binary mergers and across mass scales'' (grant No. 884631); the European Horizon Europe staff exchange (SE) programme HORIZON-MSCA-2021-SE-01 grant NewFunFiCO (No. 10108625); the Horizon ERC Grants 2021 programme under grant agreement No. 101040021; the FAPESP (Funda\c{c}\~ao de Amparo \'a Pesquisa do Estado de S\~ao Paulo) under grant 2021/01183-8; the Fondes de Recherche Nature et Technologies (FRQNT); the Fondo CAS-ANID folio CAS220010; the Generalitat Valenciana (grants APOSTD/2018/177 and  ASFAE/2022/018) and
GenT Program (project CIDEGENT/2018/021);
the Hellenic Foundation for Research and Innovation (ELIDEK) under Grant No. 23698; 
the Gordon and Betty Moore Foundation (GBMF-3561, GBMF-5278, GBMF-10423);   
the Institute for Advanced Study; the ICSC – Centro Nazionale di Ricerca in High Performance Computing, Big Data and Quantum Computing, funded by European Union – NextGenerationEU; the Istituto Nazionale di Fisica
Nucleare (INFN) sezione di Napoli, iniziative specifiche
TEONGRAV; the Research Foundation -- Flanders (FWO) Postdoctoral Fellowship 1255226N; 
the International Max Planck Research
School for Astronomy and Astrophysics at the
Universities of Bonn and Cologne; the Italian Ministry of University and Research (MUR)– Project CUP F53D23001260001, funded by the European Union – NextGenerationEU; 
Deutsche Forschungsgemeinschaft (DFG, German Research Foundation) as part of the DFG Research Unit FOR5195 -– project number 443220636;
Joint Columbia/Flatiron Postdoctoral Fellowship (research at the Flatiron Institute is supported by the Simons Foundation); 
the Japan Ministry of Education, Culture, Sports, Science and Technology (MEXT; grant JPMXP1020200109); 
the Japan Society for the Promotion of Science (JSPS) Grant-in-Aid for JSPS
Research Fellowship (JP17J08829); the Joint Institute for Computational Fundamental Science, Japan; the Key Research
Program of Frontier Sciences, Chinese Academy of
Sciences (CAS, grants QYZDJ-SSW-SLH057, QYZDJSSW-SYS008, ZDBS-LY-SLH011); 
the Leverhulme Trust Early Career Research
Fellowship; the Max-Planck-Gesellschaft (MPG);
the Max Planck Partner Group of the MPG and the
CAS; the MEXT/JSPS KAKENHI (grants 18KK0090, JP21H01137,
JP18H03721, JP18K13594, 18K03709, JP19K14761, 18H01245, 25120007, 19H01943, 21H01137, 21H04488, 22H00157, 23K03453, 24KJ0773); the MICINN Research Projects PID2019-108995GB-C22, PID2022-140888NB-C22; the MIT International Science
and Technology Initiatives (MISTI) Funds; 
the Ministry of Science and Technology (MOST) of Taiwan (103-2119-M-001-010-MY2, 105-2112-M-001-025-MY3, 105-2119-M-001-042, 106-2112-M-001-011, 106-2119-M-001-013, 106-2119-M-001-027, 106-2923-M-001-005, 107-2119-M-001-017, 107-2119-M-001-020, 107-2119-M-001-041, 107-2119-M-110-005, 107-2923-M-001-009, 108-2112-M-001-048, 108-2112-M-001-051, 108-2923-M-001-002, 109-2112-M-001-025, 109-2124-M-001-005, 109-2923-M-001-001, 
110-2112-M-001-033, 110-2124-M-001-007 and 110-2923-M-001-001); the National Science and Technology Council (NSTC) of Taiwan
(111-2124-M-001-005, 112-2124-M-001-014,  112-2112-M-003-010-MY3, and 113-2124-M-001-008);
the Ministry of Education (MoE) of Taiwan Yushan Young Scholar Program;
the Physics Division, National Center for Theoretical Sciences of Taiwan;
the National Aeronautics and
Space Administration (NASA, Fermi Guest Investigator
grant 
80NSSC23K1508, NASA Astrophysics Theory Program grant 80NSSC20K0527, NASA NuSTAR award 
80NSSC20K0645); NASA Hubble Fellowship Program Einstein Fellowship;
NASA Hubble Fellowship 
grants HST-HF2-51431.001-A, HST-HF2-51482.001-A, HST-HF2-51539.001-A, HST-HF2-51552.001A awarded 
by the Space Telescope Science Institute, which is operated by the Association of Universities for 
Research in Astronomy, Inc., for NASA, under contract NAS5-26555; 
the National Institute of Natural Sciences (NINS) of Japan; the National
Key Research and Development Program of China
(grant 2016YFA0400704, 2017YFA0402703, 2016YFA0400702); the National Science and Technology Council (NSTC, grants NSTC 111-2112-M-001 -041, NSTC 111-2124-M-001-005, NSTC 112-2124-M-001-014); the US National Science Foundation (NSF, grants AST-0096454,
AST-0352953, AST-0521233, AST-0705062, AST-0905844, AST-0922984, AST-1126433, OIA-1126433, AST-1140030,
DGE-1144085, AST-1207704, AST-1207730, AST-1207752, MRI-1228509, OPP-1248097, AST-1310896, AST-1440254, 
AST-1555365, AST-1614868, AST-1615796, AST-1715061, AST-1716327,  AST-1726637, 
OISE-1743747, AST-1743747, AST-1816420, AST-1935980, AST-1952099, AST-2034306,  AST-2205908, AST-2307887, AST-2407810, AST-2535855); 
NSF Astronomy and Astrophysics Postdoctoral Fellowship (AST-1903847); 
the NSF Graduate Research Fellowship (DGE 2140743);
the Natural Science Foundation of China (grants 11650110427, 10625314, 11721303, 11725312, 11873028, 11933007, 11991052, 11991053, 12192220, 12192223, 12273022, 12325302, 12303021); 
the AWS Impact Computing Project at the Harvard Data Science Initiative (award no. A61166); 
the Natural Sciences and Engineering Research Council of
Canada (NSERC);
the Korea Aerospace Administration (KASA) (grant no. RS-2026-25587698); 
the National Research Foundation of Korea (the Global PhD Fellowship Grant: grants NRF-2015H1A2A1033752; the Korea Research Fellowship Program: NRF-2015H1D3A1066561; Brain Pool Program: RS-2024-00407499;  Basic Research Support Grant 2019R1F1A1059721, 2021R1A6A3A01086420, 2022R1C1C1005255, RS-2022-NR071771, RS-2025-16067786, RS-2025-02214038); the POSCO Science Fellowship of the POSCO TJ Park Foundation; the Global University 30 Project Fund of Kyungpook National University in 2026; the Global-Learning \& Academic research institution for Master's \textperiodcentered~PhD students, and Postdocs(G-LAMP) Program of the National Research Foundation of Korea (NRF) grant funded by the Ministry of Education (grants RS-2023-00301914, RS-2025-25442355); NOIRLab, which is managed by the Association of Universities for Research in Astronomy (AURA) under a cooperative agreement with the National Science Foundation; 
the A.G. Leventis Foundation; 
Onsala Space Observatory (OSO) national infrastructure, for the provisioning
of its facilities/observational support (OSO receives funding through the Swedish Research Council under grant 2017-00648);  the Perimeter Institute for Theoretical Physics (research at Perimeter Institute is supported by the Government of Canada through the Department of Innovation, Science and Economic Development and by the Province of Ontario through the Ministry of Research, Innovation and Science); 
the Portuguese Foundation for Science and Technology (FCT, https://ror.org/00snfqn58) grants (Individual CEEC program – 5th edition,  FCT Multi-Annual Financing Program for R\&D Units Grants UID/04106/2025 (https://doi.org/10.54499/UID/04106/2025) and UID/PRR/04106/2025 (https://doi.org/10.54499/UID/PRR/04106/2025), PTDC/FISAST/3041/2020, CERN/FIS-PAR/0024/2021, 2022.04560.PTDC);
the Princeton Gravity Initiative; the Spanish Ministerio de Ciencia, Innovaci\'{o}n  y Universidades (grants PID2022-140888NB-C21, PID2022-140888NB-C22, PID2023-147883NB-C21, RYC2023-042988-I); the Severo Ochoa grant CEX2021-001131-S funded by MICIU/AEI/10.13039/501100011033; The European Union’s Horizon Europe research and innovation program under grant agreement No. 101093934 (RADIOBLOCKS); The European Union “NextGenerationEU”, the Recovery, Transformation and Resilience Plan, the CUII of the Andalusian Regional Government and the Spanish CSIC through grant AST22\_00001\_Subproject\_10; ``la Caixa'' Foundation (ID 100010434) through fellowship codes LCF/BQ/DI22/11940027 and LCF/BQ/DI22/11940030 and ; 
the University of Pretoria for financial aid in the provision of the new 
Cluster Server nodes and SuperMicro (USA) for a SEEDING GRANT approved toward these 
nodes in 2020; the Shanghai Municipality orientation program of basic research for international scientists (grant no. 22JC1410600); 
the Shanghai Pilot Program for Basic Research, Chinese Academy of Science, 
Shanghai Branch (JCYJ-SHFY-2021-013); the Simons Foundation (grant 00001470); the Spanish Ministry for Science and Innovation grant CEX2021-001131-S funded by MCIN/AEI/10.13039/501100011033; the Spinoza Prize SPI 78-409; the South African Research Chairs Initiative, through the 
South African Radio Astronomy Observatory (SARAO, grant ID 77948),  which is a facility of the National 
Research Foundation (NRF), an agency of the Department of Science and Innovation (DSI) of South Africa; the Swedish Research Council (VR); the Taplin Fellowship; the Toray Science Foundation; the UK Science and Technology Facilities Council (grant no. ST/X508329/1); the US Department of Energy (USDOE) through the Los Alamos National
Laboratory (operated by Triad National Security,
LLC, for the National Nuclear Security Administration
of the USDOE, contract 89233218CNA000001); and the YCAA Prize Postdoctoral Fellowship. This work was also supported by the National Research Foundation of Korea (NRF) grant funded by the Korea government(MSIT) (RS-2024-00449206). We acknowledge support from the Coordenação de Aperfeiçoamento de Pessoal de Nível Superior (CAPES) of Brazil through PROEX grant number 88887.845378/2023-00. We acknowledge financial support from Millenium Nucleus NCN23\_002 (TITANs) and Comité Mixto ESO-Chile.

We thank
the staff at the participating observatories, correlation
centers, and institutions for their enthusiastic support.
This paper makes use of the following ALMA data:
ADS/JAO.ALMA\#2017.1.00841.V and ADS/JAO.ALMA\#2019.1.01797.V.
ALMA is a partnership
of the European Southern Observatory (ESO;
Europe, representing its member states), NSF, and
National Institutes of Natural Sciences of Japan, together
with National Research Council (Canada), Ministry
of Science and Technology (MOST; Taiwan),
Academia Sinica Institute of Astronomy and Astrophysics
(ASIAA; Taiwan), and Korea Astronomy and
Space Science Institute (KASI; Republic of Korea), in
cooperation with the Republic of Chile. The Joint
ALMA Observatory is operated by ESO, Associated
Universities, Inc. (AUI)/NRAO, and the National Astronomical
Observatory of Japan (NAOJ).
The National Radio Astronomy Observatory (NRAO) and Green Bank Observatory are facilities of the U.S. National Science Foundation (NSF) operated under cooperative agreement by AUI.
This research used resources of the Oak Ridge Leadership Computing Facility at the Oak Ridge National
Laboratory, which is supported by the Office of Science of the U.S. Department of Energy under contract
No. DE-AC05-00OR22725; the ASTROVIVES FEDER infrastructure, with project code IDIFEDER-2021-086; the computing cluster of Shanghai VLBI correlator supported by the Special Fund 
for Astronomy from the Ministry of Finance in China;  
We also thank the Center for Computational Astrophysics, National Astronomical Observatory of Japan. This work was supported by FAPESP (Fundacao de Amparo a Pesquisa do Estado de Sao Paulo) under grant 2021/01183-8.

APEX is a collaboration between the
Max-Planck-Institut f{\"u}r Radioastronomie (Germany),
ESO, and the Onsala Space Observatory (Sweden). The
SMA is a joint project between the SAO and ASIAA
and is funded by the Smithsonian Institution and the
Academia Sinica. The JCMT is operated by the East
Asian Observatory on behalf of the NAOJ, ASIAA, and
KASI, as well as the Ministry of Finance of China, Chinese
Academy of Sciences, and the National Key Research and Development
Program (No. 2017YFA0402700) of China
and Natural Science Foundation of China grant 11873028.
Additional funding support for the JCMT is provided by the Science
and Technologies Facility Council (UK) and participating
universities in the UK and Canada. 
The LMT is a project operated by the Instituto Nacional
de Astr\'{o}fisica, \'{O}ptica, y Electr\'{o}nica (Mexico) and the
University of Massachusetts at Amherst (USA).
The IRAM 30 m telescope on Pico Veleta, Spain and the NOEMA interferometer on Plateau de Bure,
France are operated by IRAM and supported by CNRS (Centre National de la Recherche Scientifique, France), MPG (Max-Planck-Gesellschaft, Germany), and IGN (Instituto Geográfico Nacional, Spain).
The SMT is operated by the Arizona
Radio Observatory, a part of the Steward Observatory
of the University of Arizona, with financial support of
operations from the State of Arizona and financial support
for instrumentation development from the NSF.
Support for SPT participation in the EHT is provided by the National Science Foundation through award OPP-1852617 
to the University of Chicago. Partial support is also 
provided by the Kavli Institute of Cosmological Physics at the University of Chicago. The SPT hydrogen maser was 
provided on loan from the GLT, courtesy of ASIAA.

This work used the
Extreme Science and Engineering Discovery Environment
(XSEDE), supported by NSF grant ACI-1548562,
and CyVerse, supported by NSF grants DBI-0735191,
DBI-1265383, and DBI-1743442. XSEDE Stampede2 resource
at TACC was allocated through TG-AST170024
and TG-AST080026N. XSEDE JetStream resource at
PTI and TACC was allocated through AST170028.
This research is part of the Frontera computing project at the Texas Advanced 
Computing Center through the Frontera Large-Scale Community Partnerships allocation
AST20023. Frontera is made possible by National Science Foundation award OAC-1818253.
This research was done using services provided by the OSG Consortium 
which is supported by the National Science Foundation award Nos. 2030508 and 1836650.
Additional work used ABACUS2.0, which is part of the eScience center at Southern Denmark University, and the Kultrun Astronomy Hybrid Cluster (projects Conicyt Programa de Astronomia Fondo Quimal QUIMAL170001, Conicyt PIA ACT172033, Fondecyt Iniciacion 11170268, Quimal 220002). 
Simulations were also performed on the SuperMUC cluster at the LRZ in Garching, 
on the LOEWE cluster in CSC in Frankfurt, on the HazelHen cluster at the HLRS in Stuttgart, 
and on the Pi2.0 and Siyuan Mark-I at Shanghai Jiao Tong University.
The computer resources of the Finnish IT Center for Science (CSC) and the Finnish Computing 
Competence Infrastructure (FCCI) project are acknowledged. This
research was enabled in part by support provided
by Compute Ontario (http://computeontario.ca), Calcul
Quebec (http://www.calculquebec.ca), and the Digital Research Alliance of Canada (https://alliancecan.ca/en).

The EHTC has
received generous donations of FPGA chips from Xilinx
Inc., under the Xilinx University Program. The EHTC
has benefited from technology shared under open-source
license by the Collaboration for Astronomy Signal Processing
and Electronics Research (CASPER). The EHT
project is grateful to T4Science and Microsemi for their
assistance with hydrogen masers. This research has
made use of NASA's Astrophysics Data System. We
gratefully acknowledge the support provided by the extended
staff of the ALMA, from the inception of
the ALMA Phasing Project through the observational
campaigns of 2017 and 2018. We would like to thank
A. Deller and W. Brisken for EHT-specific support with
the use of DiFX. We thank Martin Shepherd for the addition of extra features in the Difmap software 
that were used for the CLEAN imaging results presented in this paper.
We acknowledge the significance that
Maunakea, where the SMA and JCMT EHT stations
are located, has for the indigenous Hawaiian people.
\newline

\bibliographystyle{apj_mod}
{\bibliography{bibliography}}
\clearpage
\begingroup
\providecommand{\cntext}[1]{}
\small
\setlength{\parindent}{0pt}
\setlength{\parskip}{0pt}
\raggedright
\section*{Event Horizon Telescope Collaboration: Authors and Affiliations}
\begingroup
\footnotesize
\noindent\textbf{Kazunori Akiyama} \hfill {\scriptsize ORCID: \texttt{0000-0002-9475-4254}}\par
{\footnotesize Institute of Sensors, Signals and Systems, Heriot-Watt University, Edinburgh EH14 4AS, United Kingdom}\par
{\footnotesize Massachusetts Institute of Technology Haystack Observatory, 99 Millstone Road, Westford, MA 01886, USA}\par
{\footnotesize Mizusawa VLBI Observatory, National Astronomical Observatory of Japan, 2-12 Hoshigaoka, Mizusawa, Oshu, Iwate 023-0861, Japan}\par
{\footnotesize Center for Astrophysics $|$ Harvard \& Smithsonian, 60 Garden Street, Cambridge, MA 02138, USA}\par
\vspace{0.55em}
\noindent\textbf{Ezequiel Albentosa-Ruíz} \hfill {\scriptsize ORCID: \texttt{0000-0002-7816-6401}}\par
{\footnotesize Departament d'Astronomia i Astrofísica, Universitat de València, C. Dr. Moliner 50, E-46100 Burjassot, València, Spain}\par
\vspace{0.55em}
\noindent\textbf{Antxon Alberdi} \hfill {\scriptsize ORCID: \texttt{0000-0002-9371-1033}}\par
{\footnotesize Instituto de Astrofísica de Andalucía-CSIC, Glorieta de la Astronomía s/n, E-18008 Granada, Spain}\par
\vspace{0.55em}
\noindent\textbf{Walter Alef}\par
{\footnotesize Max-Planck-Institut für Radioastronomie, Auf dem Hügel 69, D-53121 Bonn, Germany}\par
\vspace{0.55em}
\noindent\textbf{Juan Carlos Algaba} \hfill {\scriptsize ORCID: \texttt{0000-0001-6993-1696}}\par
{\footnotesize Centre for Astronomy and Astrophysics Research, Department of Physics, Faculty of Science, Universiti Malaya, 50603 Kuala Lumpur, Malaysia}\par
\vspace{0.55em}
\noindent\textbf{Rohan Ganesh Amanaganti} \hfill {\scriptsize ORCID: \texttt{0009-0002-3431-4625}}\par
{\footnotesize Center for Gravitation, Cosmology and Astrophysics, Department of Physics, University of Wisconsin–Milwaukee, P.O. Box 413, Milwaukee, WI 53201, USA}\par
\vspace{0.55em}
\noindent\textbf{Richard Anantua} \hfill {\scriptsize ORCID: \texttt{0000-0003-3457-7660}}\par
{\footnotesize Department of Physics \& Astronomy, The University of Texas at San Antonio, One UTSA Circle, San Antonio, TX 78249, USA}\par
{\footnotesize Physics \& Astronomy Department, Rice University, Houston, TX 77005-1827, USA}\par
{\footnotesize Black Hole Initiative at Harvard University, 20 Garden Street, Cambridge, MA 02138, USA}\par
{\footnotesize Center for Astrophysics $|$ Harvard \& Smithsonian, 60 Garden Street, Cambridge, MA 02138, USA}\par
\vspace{0.55em}
\noindent\textbf{Eleni Antonopoulou} \hfill {\scriptsize ORCID: \texttt{0009-0004-7747-7760}}\par
{\footnotesize Research Center for Astronomy, Academy of Athens, Soranou Efessiou 4, 115 27 Athens, Greece}\par
{\footnotesize Department of Physics, National and Kapodistrian University of Athens, Panepistimiopolis, GR 15783 Zografos, Greece}\par
\vspace{0.55em}
\noindent\textbf{Keiichi Asada} \hfill {\scriptsize ORCID: \texttt{0000-0001-6988-8763}}\par
{\footnotesize Institute of Astronomy and Astrophysics, Academia Sinica, 11F of Astronomy-Mathematics Building, AS/NTU No. 1, Sec. 4, Roosevelt Rd., Taipei 106216, Taiwan, R.O.C.}\par
\vspace{0.55em}
\noindent\textbf{Rebecca Azulay} \hfill {\scriptsize ORCID: \texttt{0000-0002-2200-5393}}\par
{\footnotesize Departament d'Astronomia i Astrofísica, Universitat de València, C. Dr. Moliner 50, E-46100 Burjassot, València, Spain}\par
{\footnotesize Observatori Astronòmic, Universitat de València, C. Catedrático José Beltrán 2, E-46980 Paterna, València, Spain}\par
{\footnotesize Max-Planck-Institut für Radioastronomie, Auf dem Hügel 69, D-53121 Bonn, Germany}\par
\vspace{0.55em}
\noindent\textbf{Anne-Kathrin Baczko} \hfill {\scriptsize ORCID: \texttt{0000-0003-3090-3975}}\par
{\footnotesize Department of Physics and Astronomy, Chalmers University of Technology, Onsala Space Observatory, SE-439 92 Onsala, Sweden}\par
{\footnotesize Max-Planck-Institut für Radioastronomie, Auf dem Hügel 69, D-53121 Bonn, Germany}\par
\vspace{0.55em}
\noindent\textbf{David Ball}\par
{\footnotesize Steward Observatory and Department of Astronomy, University of Arizona, 933 N. Cherry Ave., Tucson, AZ 85721, USA}\par
\vspace{0.55em}
\noindent\textbf{Bidisha Bandyopadhyay} \hfill {\scriptsize ORCID: \texttt{0000-0002-2138-8564}}\par
{\footnotesize Astronomy Department, Universidad de Concepción, Casilla 160-C, Concepción, Chile}\par
\vspace{0.55em}
\noindent\textbf{John Barrett} \hfill {\scriptsize ORCID: \texttt{0000-0002-9290-0764}}\par
{\footnotesize Massachusetts Institute of Technology Haystack Observatory, 99 Millstone Road, Westford, MA 01886, USA}\par
\vspace{0.55em}
\noindent\textbf{Michi Bauböck} \hfill {\scriptsize ORCID: \texttt{0000-0002-5518-2812}}\par
{\footnotesize Department of Physics, University of Illinois, 1110 West Green Street, Urbana, IL 61801, USA}\par
\vspace{0.55em}
\noindent\textbf{Bradford A. Benson} \hfill {\scriptsize ORCID: \texttt{0000-0002-5108-6823}}\par
{\footnotesize Fermi National Accelerator Laboratory, MS209, P.O. Box 500, Batavia, IL 60510, USA}\par
{\footnotesize Department of Astronomy and Astrophysics, University of Chicago, 5640 South Ellis Avenue, Chicago, IL 60637, USA}\par
\vspace{0.55em}
\noindent\textbf{Dan Bintley}\par
{\footnotesize East Asian Observatory, 660 N. A'ohoku Place, Hilo, HI 96720, USA}\par
{\footnotesize James Clerk Maxwell Telescope (JCMT), 660 N. A'ohoku Place, Hilo, HI 96720, USA}\par
\vspace{0.55em}
\noindent\textbf{Lindy Blackburn} \hfill {\scriptsize ORCID: \texttt{0000-0002-9030-642X}}\par
{\footnotesize Center for Astrophysics $|$ Harvard \& Smithsonian, 60 Garden Street, Cambridge, MA 02138, USA}\par
{\footnotesize Black Hole Initiative at Harvard University, 20 Garden Street, Cambridge, MA 02138, USA}\par
\vspace{0.55em}
\noindent\textbf{Raymond Blundell} \hfill {\scriptsize ORCID: \texttt{0000-0002-5929-5857}}\par
{\footnotesize Center for Astrophysics $|$ Harvard \& Smithsonian, 60 Garden Street, Cambridge, MA 02138, USA}\par
\vspace{0.55em}
\noindent\textbf{Katherine L. Bouman} \hfill {\scriptsize ORCID: \texttt{0000-0003-0077-4367}}\par
{\footnotesize California Institute of Technology, 1200 East California Boulevard, Pasadena, CA 91125, USA}\par
\vspace{0.55em}
\noindent\textbf{Geoffrey C. Bower} \hfill {\scriptsize ORCID: \texttt{0000-0003-4056-9982}}\par
{\footnotesize East Asian Observatory, 660 N. A'ohoku Place, Hilo, HI 96720, USA}\par
{\footnotesize James Clerk Maxwell Telescope (JCMT), 660 N. A'ohoku Place, Hilo, HI 96720, USA}\par
{\footnotesize Institute of Astronomy and Astrophysics, Academia Sinica, 645 N. A'ohoku Place, Hilo, HI 96720, USA}\par
{\footnotesize Department of Physics and Astronomy, University of Hawaii at Manoa, 2505 Correa Road, Honolulu, HI 96822, USA}\par
\vspace{0.55em}
\noindent\textbf{Michael Bremer} \hfill {\scriptsize ORCID: \texttt{0000-0001-7511-3745}}\par
{\footnotesize Institut de Radioastronomie Millimétrique (IRAM), 300 rue de la Piscine, F-38400 Saint-Martin-d'Hères, France}\par
\vspace{0.55em}
\noindent\textbf{Roger Brissenden} \hfill {\scriptsize ORCID: \texttt{0000-0002-2556-0894}}\par
{\footnotesize Center for Astrophysics $|$ Harvard \& Smithsonian, 60 Garden Street, Cambridge, MA 02138, USA}\par
\vspace{0.55em}
\noindent\textbf{Silke Britzen} \hfill {\scriptsize ORCID: \texttt{0000-0001-9240-6734}}\par
{\footnotesize Max-Planck-Institut für Radioastronomie, Auf dem Hügel 69, D-53121 Bonn, Germany}\par
\vspace{0.55em}
\noindent\textbf{Avery E. Broderick} \hfill {\scriptsize ORCID: \texttt{0000-0002-3351-760X}}\par
{\footnotesize Perimeter Institute for Theoretical Physics, 31 Caroline Street North, Waterloo, ON N2L 2Y5, Canada}\par
{\footnotesize Department of Physics and Astronomy, University of Waterloo, 200 University Avenue West, Waterloo, ON N2L 3G1, Canada}\par
{\footnotesize Waterloo Centre for Astrophysics, University of Waterloo, Waterloo, ON N2L 3G1, Canada}\par
\vspace{0.55em}
\noindent\textbf{Dominique Broguiere} \hfill {\scriptsize ORCID: \texttt{0000-0001-9151-6683}}\par
{\footnotesize Institut de Radioastronomie Millimétrique (IRAM), 300 rue de la Piscine, F-38400 Saint-Martin-d'Hères, France}\par
\vspace{0.55em}
\noindent\textbf{Thomas Bronzwaer} \hfill {\scriptsize ORCID: \texttt{0000-0003-1151-3971}}\par
{\footnotesize Department of Astrophysics, Institute for Mathematics, Astrophysics and Particle Physics (IMAPP), Radboud University, P.O. Box 9010, 6500 GL Nijmegen, The Netherlands}\par
\vspace{0.55em}
\noindent\textbf{Sandra Bustamante} \hfill {\scriptsize ORCID: \texttt{0000-0001-6169-1894}}\par
{\footnotesize Department of Astronomy, University of Massachusetts, Amherst, MA 01003, USA}\par
\vspace{0.55em}
\noindent\textbf{Douglas F. Carlos} \hfill {\scriptsize ORCID: \texttt{0000-0002-1340-7702}}\par
{\footnotesize Instituto de Astronomia, Geofísica e Ciências Atmosféricas, Universidade de São Paulo, R. do Matão, 1226, São Paulo, SP 05508-090, Brazil}\par
\vspace{0.55em}
\noindent\textbf{John E. Carlstrom} \hfill {\scriptsize ORCID: \texttt{0000-0002-2044-7665}}\par
{\footnotesize Kavli Institute for Cosmological Physics, University of Chicago, 5640 South Ellis Avenue, Chicago, IL 60637, USA}\par
{\footnotesize Department of Astronomy and Astrophysics, University of Chicago, 5640 South Ellis Avenue, Chicago, IL 60637, USA}\par
{\footnotesize Department of Physics, University of Chicago, 5720 South Ellis Avenue, Chicago, IL 60637, USA}\par
{\footnotesize Enrico Fermi Institute, University of Chicago, 5640 South Ellis Avenue, Chicago, IL 60637, USA}\par
\vspace{0.55em}
\noindent\textbf{Andrew Chael} \hfill {\scriptsize ORCID: \texttt{0000-0003-2966-6220}}\par
{\footnotesize Princeton Gravity Initiative, Jadwin Hall, Princeton University, Princeton, NJ 08544, USA}\par
\vspace{0.55em}
\noindent\textbf{Chi-kwan Chan} \hfill {\scriptsize ORCID: \texttt{0000-0001-6337-6126}}\par
{\footnotesize Steward Observatory and Department of Astronomy, University of Arizona, 933 N. Cherry Ave., Tucson, AZ 85721, USA}\par
{\footnotesize Data Science Institute, University of Arizona, 1230 N. Cherry Ave., Tucson, AZ 85721, USA}\par
{\footnotesize Program in Applied Mathematics, University of Arizona, 617 N. Santa Rita, Tucson, AZ 85721, USA}\par
\vspace{0.55em}
\noindent\textbf{Chin-Shin Chang} \hfill {\scriptsize ORCID: \texttt{0000-0001-9910-3234}}\par
{\footnotesize Department of Astronomy, University of Geneva, Chemin Pegasi 51, 1290 Versoix, Switzerland}\par
\vspace{0.55em}
\noindent\textbf{Dominic O. Chang} \hfill {\scriptsize ORCID: \texttt{0000-0001-9939-5257}}\par
{\footnotesize Center for Astrophysics $|$ Harvard \& Smithsonian, 60 Garden Street, Cambridge, MA 02138, USA}\par
{\footnotesize Black Hole Initiative at Harvard University, 20 Garden Street, Cambridge, MA 02138, USA}\par
\vspace{0.55em}
\noindent\textbf{Koushik Chatterjee} \hfill {\scriptsize ORCID: \texttt{0000-0002-2825-3590}}\par
{\footnotesize Department of Physics, University of Maryland, 7901 Regents Drive, College Park, MD 20742, USA}\par
{\footnotesize Black Hole Initiative at Harvard University, 20 Garden Street, Cambridge, MA 02138, USA}\par
{\footnotesize Center for Astrophysics $|$ Harvard \& Smithsonian, 60 Garden Street, Cambridge, MA 02138, USA}\par
\vspace{0.55em}
\noindent\textbf{Erandi Chavez} \hfill {\scriptsize ORCID: \texttt{0000-0003-4143-9717}}\par
{\footnotesize Center for Astrophysics $|$ Harvard \& Smithsonian, 60 Garden Street, Cambridge, MA 02138, USA}\par
\vspace{0.55em}
\noindent\textbf{Ming-Tang Chen} \hfill {\scriptsize ORCID: \texttt{0000-0001-6573-3318}}\par
{\footnotesize Institute of Astronomy and Astrophysics, Academia Sinica, 645 N. A'ohoku Place, Hilo, HI 96720, USA}\par
\vspace{0.55em}
\noindent\textbf{Yongjun Chen (\cntext{陈永军})} \hfill {\scriptsize ORCID: \texttt{0000-0001-5650-6770}}\par
{\footnotesize Shanghai Astronomical Observatory, Chinese Academy of Sciences, 80 Nandan Road, Shanghai 200030, People's Republic of China}\par
{\footnotesize Key Laboratory of Radio Astronomy and Technology, Chinese Academy of Sciences, A20 Datun Road, Chaoyang District, Beijing, 100101, People’s Republic of China}\par
\vspace{0.55em}
\noindent\textbf{Xiaopeng Cheng} \hfill {\scriptsize ORCID: \texttt{0000-0003-4407-9868}}\par
{\footnotesize Korea Astronomy and Space Science Institute, Daedeok-daero 776, Yuseong-gu, Daejeon 34055, Republic of Korea}\par
\vspace{0.55em}
\noindent\textbf{Paul Chichura} \hfill {\scriptsize ORCID: \texttt{0000-0002-5397-9035}}\par
{\footnotesize Department of Physics, University of Chicago, 5720 South Ellis Avenue, Chicago, IL 60637, USA}\par
{\footnotesize Kavli Institute for Cosmological Physics, University of Chicago, 5640 South Ellis Avenue, Chicago, IL 60637, USA}\par
\vspace{0.55em}
\noindent\textbf{Ilje Cho} \hfill {\scriptsize ORCID: \texttt{0000-0001-6083-7521}}\par
{\footnotesize Department of Astronomy, Kyungpook National University, 80 Daehak-ro, Buk-gu, Daegu 41566, Republic of Korea}\par
{\footnotesize Korea Astronomy and Space Science Institute, Daedeok-daero 776, Yuseong-gu, Daejeon 34055, Republic of Korea}\par
{\footnotesize Instituto de Astrofísica de Andalucía-CSIC, Glorieta de la Astronomía s/n, E-18008 Granada, Spain}\par
\vspace{0.55em}
\noindent\textbf{Nicholas S. Conroy} \hfill {\scriptsize ORCID: \texttt{0000-0003-2886-2377}}\par
{\footnotesize Department of Astronomy, University of Illinois at Urbana-Champaign, 1002 West Green Street, Urbana, IL 61801, USA}\par
{\footnotesize Center for Astrophysics $|$ Harvard \& Smithsonian, 60 Garden Street, Cambridge, MA 02138, USA}\par
\vspace{0.55em}
\noindent\textbf{John E. Conway} \hfill {\scriptsize ORCID: \texttt{0000-0003-2448-9181}}\par
{\footnotesize Department of Physics and Astronomy, Chalmers University of Technology, Onsala Space Observatory, SE-439 92 Onsala, Sweden}\par
\vspace{0.55em}
\noindent\textbf{Thomas M. Crawford} \hfill {\scriptsize ORCID: \texttt{0000-0001-9000-5013}}\par
{\footnotesize Department of Astronomy and Astrophysics, University of Chicago, 5640 South Ellis Avenue, Chicago, IL 60637, USA}\par
{\footnotesize Kavli Institute for Cosmological Physics, University of Chicago, 5640 South Ellis Avenue, Chicago, IL 60637, USA}\par
\vspace{0.55em}
\noindent\textbf{Geoffrey B. Crew} \hfill {\scriptsize ORCID: \texttt{0000-0002-2079-3189}}\par
{\footnotesize Massachusetts Institute of Technology Haystack Observatory, 99 Millstone Road, Westford, MA 01886, USA}\par
\vspace{0.55em}
\noindent\textbf{Alejandro Cruz-Osorio} \hfill {\scriptsize ORCID: \texttt{0000-0002-3945-6342}}\par
{\footnotesize Instituto de Astronomía, Universidad Nacional Autónoma de México (UNAM), Apdo Postal 70-264, Ciudad de México, México}\par
\vspace{0.55em}
\noindent\textbf{Yuzhu Cui (\cntext{崔玉竹})} \hfill {\scriptsize ORCID: \texttt{0000-0001-6311-4345}}\par
{\footnotesize Institute of Astrophysics, Central China Normal University, Wuhan 430079, People's Republic of China}\par
\vspace{0.55em}
\noindent\textbf{Brandon Curd} \hfill {\scriptsize ORCID: \texttt{0000-0002-8650-0879}}\par
{\footnotesize Department of Physics \& Astronomy, The University of Texas at San Antonio, One UTSA Circle, San Antonio, TX 78249, USA}\par
{\footnotesize Black Hole Initiative at Harvard University, 20 Garden Street, Cambridge, MA 02138, USA}\par
{\footnotesize Center for Astrophysics $|$ Harvard \& Smithsonian, 60 Garden Street, Cambridge, MA 02138, USA}\par
\vspace{0.55em}
\noindent\textbf{Rohan Dahale} \hfill {\scriptsize ORCID: \texttt{0000-0001-6982-9034}}\par
{\footnotesize Instituto de Astrofísica de Andalucía-CSIC, Glorieta de la Astronomía s/n, E-18008 Granada, Spain}\par
{\footnotesize Department of Computer Science, University of Toronto, 40 St. George St., Toronto, ON, M5S 2E4, Canada}\par
{\footnotesize Canadian Institute for Theoretical Astrophysics, University of Toronto, 60 St. George Street, Toronto, ON M5S 3H8, Canada}\par
\vspace{0.55em}
\noindent\textbf{Jordy Davelaar} \hfill {\scriptsize ORCID: \texttt{0000-0002-2685-2434}}\par
{\footnotesize Department of Astrophysical Sciences, Peyton Hall, Princeton University, Princeton, NJ 08544, USA}\par
{\footnotesize NASA Hubble Fellowship Program, Einstein Fellow}\par
\vspace{0.55em}
\noindent\textbf{Joost de Kleuver} \hfill {\scriptsize ORCID: \texttt{0009-0001-9624-1751}}\par
{\footnotesize Department of Astrophysics, Institute for Mathematics, Astrophysics and Particle Physics (IMAPP), Radboud University, P.O. Box 9010, 6500 GL Nijmegen, The Netherlands}\par
\vspace{0.55em}
\noindent\textbf{Mariafelicia De Laurentis} \hfill {\scriptsize ORCID: \texttt{0000-0002-9945-682X}}\par
{\footnotesize Dipartimento di Fisica ``E. Pancini'', Università di Napoli ``Federico II'', Compl. Univ. di Monte S. Angelo, Edificio G, Via Cinthia, I-80126, Napoli, Italy}\par
{\footnotesize INFN Sez. di Napoli, Compl. Univ. di Monte S. Angelo, Edificio G, Via Cinthia, I-80126, Napoli, Italy}\par
\vspace{0.55em}
\noindent\textbf{Roger Deane} \hfill {\scriptsize ORCID: \texttt{0000-0003-1027-5043}}\par
{\footnotesize Wits Centre for Astrophysics, University of the Witwatersrand, 1 Jan Smuts Avenue, Braamfontein, Johannesburg 2050, South Africa}\par
{\footnotesize Department of Physics, University of Pretoria, Hatfield, Pretoria 0028, South Africa}\par
{\footnotesize Centre for Radio Astronomy Techniques and Technologies, Department of Physics and Electronics, Rhodes University, Makhanda 6140, South Africa}\par
\vspace{0.55em}
\noindent\textbf{Jason Dexter} \hfill {\scriptsize ORCID: \texttt{0000-0003-3903-0373}}\par
{\footnotesize JILA and Department of Astrophysical and Planetary Sciences, University of Colorado, Boulder, CO 80309, USA}\par
\vspace{0.55em}
\noindent\textbf{Vedant Dhruv} \hfill {\scriptsize ORCID: \texttt{0000-0001-6765-877X}}\par
{\footnotesize Department of Physics, University of Illinois, 1110 West Green Street, Urbana, IL 61801, USA}\par
\vspace{0.55em}
\noindent\textbf{Indu K. Dihingia} \hfill {\scriptsize ORCID: \texttt{0000-0002-4064-0446}}\par
{\footnotesize Institute of Fundamental Physics and Quantum Technology, \& School of Physical Science and Technology, Ningbo University, Ningbo, Zhejiang 315211, People’s Republic of China}\par
{\footnotesize Tsung-Dao Lee Institute, Shanghai Jiao Tong University, Shengrong Road 520, Shanghai, 201210, People’s Republic of China}\par
\vspace{0.55em}
\noindent\textbf{Sheperd S. Doeleman} \hfill {\scriptsize ORCID: \texttt{0000-0002-9031-0904}}\par
{\footnotesize Center for Astrophysics $|$ Harvard \& Smithsonian, 60 Garden Street, Cambridge, MA 02138, USA}\par
{\footnotesize Black Hole Initiative at Harvard University, 20 Garden Street, Cambridge, MA 02138, USA}\par
\vspace{0.55em}
\noindent\textbf{Sergio A. Dzib} \hfill {\scriptsize ORCID: \texttt{0000-0001-6010-6200}}\par
{\footnotesize Max-Planck-Institut für Radioastronomie, Auf dem Hügel 69, D-53121 Bonn, Germany}\par
\vspace{0.55em}
\noindent\textbf{Razieh Emami} \hfill {\scriptsize ORCID: \texttt{0000-0002-2791-5011}}\par
{\footnotesize School of Astronomy, Institute for Research in Fundamental Sciences (IPM) P.O. Box 19395-5746, Tehran, Iran}\par
\vspace{0.55em}
\noindent\textbf{Heino Falcke} \hfill {\scriptsize ORCID: \texttt{0000-0002-2526-6724}}\par
{\footnotesize Department of Astrophysics, Institute for Mathematics, Astrophysics and Particle Physics (IMAPP), Radboud University, P.O. Box 9010, 6500 GL Nijmegen, The Netherlands}\par
\vspace{0.55em}
\noindent\textbf{Joseph R. Farah} \hfill {\scriptsize ORCID: \texttt{0000-0003-4914-5625}}\par
{\footnotesize Miller Institute for Basic Research in Science, 206B Stanley Hall, Berkeley, CA 94720, USA}\par
{\footnotesize Department of Astronomy, University of California, Berkeley, CA 94720-3411, USA}\par
\vspace{0.55em}
\noindent\textbf{Vincent L. Fish} \hfill {\scriptsize ORCID: \texttt{0000-0002-7128-9345}}\par
{\footnotesize Massachusetts Institute of Technology Haystack Observatory, 99 Millstone Road, Westford, MA 01886, USA}\par
\vspace{0.55em}
\noindent\textbf{Edward Fomalont} \hfill {\scriptsize ORCID: \texttt{0000-0002-9036-2747}}\par
{\footnotesize National Radio Astronomy Observatory, 520 Edgemont Road, Charlottesville, VA 22903, USA}\par
\vspace{0.55em}
\noindent\textbf{H. Alyson Ford} \hfill {\scriptsize ORCID: \texttt{0000-0002-9797-0972}}\par
{\footnotesize Steward Observatory and Department of Astronomy, University of Arizona, 933 N. Cherry Ave., Tucson, AZ 85721, USA}\par
\vspace{0.55em}
\noindent\textbf{Marianna Foschi} \hfill {\scriptsize ORCID: \texttt{0000-0001-8147-4993}}\par
{\footnotesize Instituto de Astrofísica de Andalucía-CSIC, Glorieta de la Astronomía s/n, E-18008 Granada, Spain}\par
{\footnotesize California Institute of Technology, 1200 East California Boulevard, Pasadena, CA 91125, USA}\par
\vspace{0.55em}
\noindent\textbf{Raquel Fraga-Encinas} \hfill {\scriptsize ORCID: \texttt{0000-0002-5222-1361}}\par
{\footnotesize Department of Astrophysics, Institute for Mathematics, Astrophysics and Particle Physics (IMAPP), Radboud University, P.O. Box 9010, 6500 GL Nijmegen, The Netherlands}\par
\vspace{0.55em}
\noindent\textbf{William T. Freeman}\par
{\footnotesize Department of Electrical Engineering and Computer Science, Massachusetts Institute of Technology, 32-D476, 77 Massachusetts Ave., Cambridge, MA 02142, USA}\par
{\footnotesize Google Research, 355 Main St., Cambridge, MA 02142, USA}\par
\vspace{0.55em}
\noindent\textbf{Per Friberg} \hfill {\scriptsize ORCID: \texttt{0000-0002-8010-8454}}\par
{\footnotesize East Asian Observatory, 660 N. A'ohoku Place, Hilo, HI 96720, USA}\par
{\footnotesize James Clerk Maxwell Telescope (JCMT), 660 N. A'ohoku Place, Hilo, HI 96720, USA}\par
\vspace{0.55em}
\noindent\textbf{Christian M. Fromm} \hfill {\scriptsize ORCID: \texttt{0000-0002-1827-1656}}\par
{\footnotesize Institut für Theoretische Physik und Astrophysik, Universität Würzburg, Emil-Fischer-Str. 31, D-97074 Würzburg, Germany}\par
{\footnotesize Institut für Theoretische Physik, Goethe-Universität Frankfurt, Max-von-Laue-Straße 1, D-60438 Frankfurt am Main, Germany}\par
{\footnotesize Max-Planck-Institut für Radioastronomie, Auf dem Hügel 69, D-53121 Bonn, Germany}\par
\vspace{0.55em}
\noindent\textbf{Antonio Fuentes} \hfill {\scriptsize ORCID: \texttt{0000-0002-8773-4933}}\par
{\footnotesize Instituto de Astrofísica de Andalucía-CSIC, Glorieta de la Astronomía s/n, E-18008 Granada, Spain}\par
\vspace{0.55em}
\noindent\textbf{Peter Galison} \hfill {\scriptsize ORCID: \texttt{0000-0002-6429-3872}}\par
{\footnotesize Black Hole Initiative at Harvard University, 20 Garden Street, Cambridge, MA 02138, USA}\par
{\footnotesize Department of History of Science, Harvard University, Cambridge, MA 02138, USA}\par
{\footnotesize Department of Physics, Harvard University, Cambridge, MA 02138, USA}\par
\vspace{0.55em}
\noindent\textbf{Charles F. Gammie} \hfill {\scriptsize ORCID: \texttt{0000-0001-7451-8935}}\par
{\footnotesize Department of Physics, University of Illinois, 1110 West Green Street, Urbana, IL 61801, USA}\par
{\footnotesize Department of Astronomy, University of Illinois at Urbana-Champaign, 1002 West Green Street, Urbana, IL 61801, USA}\par
{\footnotesize NCSA, University of Illinois, 1205 W. Clark St., Urbana, IL 61801, USA}\par
\vspace{0.55em}
\noindent\textbf{Roberto García} \hfill {\scriptsize ORCID: \texttt{0000-0002-6584-7443}}\par
{\footnotesize Institut de Radioastronomie Millimétrique (IRAM), 300 rue de la Piscine, F-38400 Saint-Martin-d'Hères, France}\par
\vspace{0.55em}
\noindent\textbf{Olivier Gentaz} \hfill {\scriptsize ORCID: \texttt{0000-0002-0115-4605}}\par
{\footnotesize Institut de Radioastronomie Millimétrique (IRAM), 300 rue de la Piscine, F-38400 Saint-Martin-d'Hères, France}\par
\vspace{0.55em}
\noindent\textbf{Boris Georgiev} \hfill {\scriptsize ORCID: \texttt{0000-0002-3586-6424}}\par
{\footnotesize Steward Observatory and Department of Astronomy, University of Arizona, 933 N. Cherry Ave., Tucson, AZ 85721, USA}\par
\vspace{0.55em}
\noindent\textbf{Ciriaco Goddi} \hfill {\scriptsize ORCID: \texttt{0000-0002-2542-7743}}\par
{\footnotesize Instituto de Astronomia, Geofísica e Ciências Atmosféricas, Universidade de São Paulo, R. do Matão, 1226, São Paulo, SP 05508-090, Brazil}\par
{\footnotesize Dipartimento di Fisica, Università degli Studi di Cagliari, SP Monserrato-Sestu km 0.7, I-09042 Monserrato (CA), Italy}\par
{\footnotesize INAF - Osservatorio Astronomico di Cagliari, via della Scienza 5, I-09047 Selargius (CA), Italy}\par
{\footnotesize INFN, sezione di Cagliari, I-09042 Monserrato (CA), Italy}\par
\vspace{0.55em}
\noindent\textbf{Roman Gold} \hfill {\scriptsize ORCID: \texttt{0000-0003-2492-1966}}\par
{\footnotesize Institute for Mathematics and Interdisciplinary Center for Scientific Computing, Heidelberg University, Im Neuenheimer Feld 205, Heidelberg 69120, Germany}\par
{\footnotesize Institut f\"ur Theoretische Physik, Universit\"at Heidelberg, Philosophenweg 16, 69120 Heidelberg, Germany}\par
{\footnotesize CP3-Origins, University of Southern Denmark, Campusvej 55, DK-5230 Odense, Denmark}\par
\vspace{0.55em}
\noindent\textbf{Arturo I. Gómez-Ruiz} \hfill {\scriptsize ORCID: \texttt{0000-0001-9395-1670}}\par
{\footnotesize Instituto Nacional de Astrofísica, Óptica y Electrónica. Apartado Postal 51 y 216, 72000. Puebla Pue., México}\par
{\footnotesize Consejo Nacional de Humanidades, Ciencia y Tecnología, Av. Insurgentes Sur 1582, 03940, Ciudad de México, México}\par
\vspace{0.55em}
\noindent\textbf{Brissa Gomez Miller} \hfill {\scriptsize ORCID: \texttt{0009-0006-8345-6805}}\par
{\footnotesize Instituto de Radioastronomía y Astrofísica, Universidad Nacional Autónoma de México, Morelia 58089, México}\par
\vspace{0.55em}
\noindent\textbf{José L. Gómez} \hfill {\scriptsize ORCID: \texttt{0000-0003-4190-7613}}\par
{\footnotesize Instituto de Astrofísica de Andalucía-CSIC, Glorieta de la Astronomía s/n, E-18008 Granada, Spain}\par
\vspace{0.55em}
\noindent\textbf{Minfeng Gu (\cntext{顾敏峰})} \hfill {\scriptsize ORCID: \texttt{0000-0002-4455-6946}}\par
{\footnotesize Shanghai Astronomical Observatory, Chinese Academy of Sciences, 80 Nandan Road, Shanghai 200030, People's Republic of China}\par
{\footnotesize Key Laboratory for Research in Galaxies and Cosmology, Chinese Academy of Sciences, Shanghai 200030, People's Republic of China}\par
\vspace{0.55em}
\noindent\textbf{Mark Gurwell} \hfill {\scriptsize ORCID: \texttt{0000-0003-0685-3621}}\par
{\footnotesize Center for Astrophysics $|$ Harvard \& Smithsonian, 60 Garden Street, Cambridge, MA 02138, USA}\par
\vspace{0.55em}
\noindent\textbf{Kazuhiro Hada} \hfill {\scriptsize ORCID: \texttt{0000-0001-6906-772X}}\par
{\footnotesize Graduate School of Science, Nagoya City University, Yamanohata 1, Mizuho-cho, Mizuho-ku, Nagoya, 467-8501, Aichi, Japan}\par
{\footnotesize Mizusawa VLBI Observatory, National Astronomical Observatory of Japan, 2-12 Hoshigaoka, Mizusawa, Oshu, Iwate 023-0861, Japan}\par
\vspace{0.55em}
\noindent\textbf{Daryl Haggard} \hfill {\scriptsize ORCID: \texttt{0000-0001-6803-2138}}\par
{\footnotesize Department of Physics, McGill University, 3600 rue University, Montréal, QC H3A 2T8, Canada}\par
{\footnotesize Trottier Space Institute at McGill, 3550 rue University, Montréal, QC H3A 2A7, Canada}\par
\vspace{0.55em}
\noindent\textbf{Ronald Hesper} \hfill {\scriptsize ORCID: \texttt{0000-0003-1918-6098}}\par
{\footnotesize NOVA Sub-mm Instrumentation Group, Kapteyn Astronomical Institute, University of Groningen, Landleven 12, 9747 AD Groningen, The Netherlands}\par
\vspace{0.55em}
\noindent\textbf{Dirk Heumann} \hfill {\scriptsize ORCID: \texttt{0000-0002-7671-0047}}\par
{\footnotesize Steward Observatory and Department of Astronomy, University of Arizona, 933 N. Cherry Ave., Tucson, AZ 85721, USA}\par
\vspace{0.55em}
\noindent\textbf{Luis C. Ho (\cntext{何子山})} \hfill {\scriptsize ORCID: \texttt{0000-0001-6947-5846}}\par
{\footnotesize Department of Astronomy, School of Physics, Peking University, Beijing 100871, People's Republic of China}\par
{\footnotesize Kavli Institute for Astronomy and Astrophysics, Peking University, Beijing 100871, People's Republic of China}\par
\vspace{0.55em}
\noindent\textbf{Paul Ho} \hfill {\scriptsize ORCID: \texttt{0000-0002-3412-4306}}\par
{\footnotesize Institute of Astronomy and Astrophysics, Academia Sinica, 11F of Astronomy-Mathematics Building, AS/NTU No. 1, Sec. 4, Roosevelt Rd., Taipei 106216, Taiwan, R.O.C.}\par
{\footnotesize James Clerk Maxwell Telescope (JCMT), 660 N. A'ohoku Place, Hilo, HI 96720, USA}\par
{\footnotesize East Asian Observatory, 660 N. A'ohoku Place, Hilo, HI 96720, USA}\par
\vspace{0.55em}
\noindent\textbf{Daniel Hoak} \hfill {\scriptsize ORCID: \texttt{0000-0001-5126-8048}}\par
{\footnotesize Massachusetts Institute of Technology Haystack Observatory, 99 Millstone Road, Westford, MA 01886, USA}\par
\vspace{0.55em}
\noindent\textbf{Mareki Honma} \hfill {\scriptsize ORCID: \texttt{0000-0003-4058-9000}}\par
{\footnotesize Mizusawa VLBI Observatory, National Astronomical Observatory of Japan, 2-12 Hoshigaoka, Mizusawa, Oshu, Iwate 023-0861, Japan}\par
{\footnotesize Department of Astronomical Science, The Graduate University for Advanced Studies (SOKENDAI), 2-21-1 Osawa, Mitaka, Tokyo 181-8588, Japan}\par
{\footnotesize Department of Astronomy, Graduate School of Science, The University of Tokyo, 7-3-1 Hongo, Bunkyo-ku, Tokyo 113-0033, Japan}\par
\vspace{0.55em}
\noindent\textbf{Chih-Wei L. Huang} \hfill {\scriptsize ORCID: \texttt{0000-0001-5641-3953}}\par
{\footnotesize Institute of Astronomy and Astrophysics, Academia Sinica, 11F of Astronomy-Mathematics Building, AS/NTU No. 1, Sec. 4, Roosevelt Rd., Taipei 106216, Taiwan, R.O.C.}\par
\vspace{0.55em}
\noindent\textbf{Lei Huang (\cntext{黄磊})} \hfill {\scriptsize ORCID: \texttt{0000-0002-1923-227X}}\par
{\footnotesize Shanghai Astronomical Observatory, Chinese Academy of Sciences, 80 Nandan Road, Shanghai 200030, People's Republic of China}\par
{\footnotesize Key Laboratory for Research in Galaxies and Cosmology, Chinese Academy of Sciences, Shanghai 200030, People's Republic of China}\par
\vspace{0.55em}
\noindent\textbf{David H. Hughes}\par
{\footnotesize Instituto Nacional de Astrofísica, Óptica y Electrónica. Apartado Postal 51 y 216, 72000. Puebla Pue., México}\par
\vspace{0.55em}
\noindent\textbf{Shiro Ikeda} \hfill {\scriptsize ORCID: \texttt{0000-0002-2462-1448}}\par
{\footnotesize National Astronomical Observatory of Japan, 2-21-1 Osawa, Mitaka, Tokyo 181-8588, Japan}\par
{\footnotesize The Institute of Statistical Mathematics, 10-3 Midori-cho, Tachikawa, Tokyo, 190-8562, Japan}\par
{\footnotesize Department of Statistical Science, The Graduate University for Advanced Studies (SOKENDAI), 10-3 Midori-cho, Tachikawa, Tokyo 190-8562, Japan}\par
{\footnotesize Kavli Institute for the Physics and Mathematics of the Universe, The University of Tokyo, 5-1-5 Kashiwanoha, Kashiwa, 277-8583, Japan}\par
\vspace{0.55em}
\noindent\textbf{C. M. Violette Impellizzeri} \hfill {\scriptsize ORCID: \texttt{0000-0002-3443-2472}}\par
{\footnotesize Leiden Observatory, Leiden University, Postbus 2300, 9513 RA Leiden, The Netherlands}\par
{\footnotesize National Radio Astronomy Observatory, 520 Edgemont Road, Charlottesville, VA 22903, USA}\par
\vspace{0.55em}
\noindent\textbf{Makoto Inoue} \hfill {\scriptsize ORCID: \texttt{0000-0001-5037-3989}}\par
{\footnotesize Institute of Astronomy and Astrophysics, Academia Sinica, 11F of Astronomy-Mathematics Building, AS/NTU No. 1, Sec. 4, Roosevelt Rd., Taipei 106216, Taiwan, R.O.C.}\par
\vspace{0.55em}
\noindent\textbf{Sara Issaoun} \hfill {\scriptsize ORCID: \texttt{0000-0002-5297-921X}}\par
{\footnotesize Center for Astrophysics $|$ Harvard \& Smithsonian, 60 Garden Street, Cambridge, MA 02138, USA}\par
{\footnotesize NASA Hubble Fellowship Program, Einstein Fellow}\par
\vspace{0.55em}
\noindent\textbf{Yuhei Iwata} \hfill {\scriptsize ORCID: \texttt{0000-0002-9255-4742}}\par
{\footnotesize Mizusawa VLBI Observatory, National Astronomical Observatory of Japan, 2-12 Hoshigaoka, Mizusawa, Oshu, Iwate 023-0861, Japan}\par
{\footnotesize Department of Astronomical Science, The Graduate University for Advanced Studies (SOKENDAI), 2-21-1 Osawa, Mitaka, Tokyo 181-8588, Japan}\par
\vspace{0.55em}
\noindent\textbf{David J. James} \hfill {\scriptsize ORCID: \texttt{0000-0001-5160-4486}}\par
{\footnotesize ASTRAVEO LLC, PO Box 1668, Gloucester, MA 01931, USA}\par
{\footnotesize Applied Materials Inc., 35 Dory Road, Gloucester, MA 01930, USA}\par
\vspace{0.55em}
\noindent\textbf{Buell T. Jannuzi} \hfill {\scriptsize ORCID: \texttt{0000-0002-1578-6582}}\par
{\footnotesize Steward Observatory and Department of Astronomy, University of Arizona, 933 N. Cherry Ave., Tucson, AZ 85721, USA}\par
\vspace{0.55em}
\noindent\textbf{Michael Janssen} \hfill {\scriptsize ORCID: \texttt{0000-0001-8685-6544}}\par
{\footnotesize Department of Astrophysics, Institute for Mathematics, Astrophysics and Particle Physics (IMAPP), Radboud University, P.O. Box 9010, 6500 GL Nijmegen, The Netherlands}\par
{\footnotesize Max-Planck-Institut für Radioastronomie, Auf dem Hügel 69, D-53121 Bonn, Germany}\par
\vspace{0.55em}
\noindent\textbf{Britton Jeter} \hfill {\scriptsize ORCID: \texttt{0000-0003-2847-1712}}\par
{\footnotesize Finnish Centre for Astronomy with ESO, University of Turku, FI-20014 Turun Yliopisto, Finland}\par
{\footnotesize Aalto University Metsähovi Radio Observatory, Metsähovintie 114, FI-02540 Kylmälä, Finland}\par
\vspace{0.55em}
\noindent\textbf{Wu Jiang (\cntext{江悟})} \hfill {\scriptsize ORCID: \texttt{0000-0001-7369-3539}}\par
{\footnotesize Shanghai Astronomical Observatory, Chinese Academy of Sciences, 80 Nandan Road, Shanghai 200030, People's Republic of China}\par
\vspace{0.55em}
\noindent\textbf{Alejandra Jiménez-Rosales} \hfill {\scriptsize ORCID: \texttt{0000-0002-2662-3754}}\par
{\footnotesize Department of Astrophysics, Institute for Mathematics, Astrophysics and Particle Physics (IMAPP), Radboud University, P.O. Box 9010, 6500 GL Nijmegen, The Netherlands}\par
\vspace{0.55em}
\noindent\textbf{Michael D. Johnson} \hfill {\scriptsize ORCID: \texttt{0000-0002-4120-3029}}\par
{\footnotesize Center for Astrophysics $|$ Harvard \& Smithsonian, 60 Garden Street, Cambridge, MA 02138, USA}\par
{\footnotesize Black Hole Initiative at Harvard University, 20 Garden Street, Cambridge, MA 02138, USA}\par
\vspace{0.55em}
\noindent\textbf{Svetlana Jorstad} \hfill {\scriptsize ORCID: \texttt{0000-0001-6158-1708}}\par
{\footnotesize Institute for Astrophysical Research, Boston University, 725 Commonwealth Ave., Boston, MA 02215, USA}\par
\vspace{0.55em}
\noindent\textbf{Adam C. Jones}\par
{\footnotesize Department of Astronomy and Astrophysics, University of Chicago, 5640 South Ellis Avenue, Chicago, IL 60637, USA}\par
\vspace{0.55em}
\noindent\textbf{Abhishek V. Joshi} \hfill {\scriptsize ORCID: \texttt{0000-0002-2514-5965}}\par
{\footnotesize Department of Physics, University of Illinois, 1110 West Green Street, Urbana, IL 61801, USA}\par
\vspace{0.55em}
\noindent\textbf{Taehyun Jung} \hfill {\scriptsize ORCID: \texttt{0000-0001-7003-8643}}\par
{\footnotesize Korea Astronomy and Space Science Institute, Daedeok-daero 776, Yuseong-gu, Daejeon 34055, Republic of Korea}\par
{\footnotesize Korea National University of Science and Technology, Gajeong-ro 217, Yuseong-gu, Daejeon 34113, Republic of Korea}\par
\vspace{0.55em}
\noindent\textbf{Tomohisa Kawashima} \hfill {\scriptsize ORCID: \texttt{0000-0001-8527-0496}}\par
{\footnotesize National Institute of Technology, Ichinoseki College, Takanashi, Hagisho, Ichinoseki, Iwate, 021-8511, Japan}\par
\vspace{0.55em}
\noindent\textbf{Garrett K. Keating} \hfill {\scriptsize ORCID: \texttt{0000-0002-3490-146X}}\par
{\footnotesize Center for Astrophysics $|$ Harvard \& Smithsonian, 60 Garden Street, Cambridge, MA 02138, USA}\par
\vspace{0.55em}
\noindent\textbf{Mark Kettenis} \hfill {\scriptsize ORCID: \texttt{0000-0002-6156-5617}}\par
{\footnotesize Joint Institute for VLBI ERIC (JIVE), Oude Hoogeveensedijk 4, 7991 PD Dwingeloo, The Netherlands}\par
\vspace{0.55em}
\noindent\textbf{Dong-Jin Kim} \hfill {\scriptsize ORCID: \texttt{0000-0002-7038-2118}}\par
{\footnotesize CSIRO, Space and Astronomy, PO Box 76, Epping, NSW 1710, Australia}\par
\vspace{0.55em}
\noindent\textbf{Jae-Young Kim} \hfill {\scriptsize ORCID: \texttt{0000-0001-8229-7183}}\par
{\footnotesize Department of Physics, Ulsan National Institute of Science and Technology (UNIST), Ulsan 44919, Republic of Korea}\par
\vspace{0.55em}
\noindent\textbf{Jongsoo Kim} \hfill {\scriptsize ORCID: \texttt{0000-0002-1229-0426}}\par
{\footnotesize Korea Astronomy and Space Science Institute, Daedeok-daero 776, Yuseong-gu, Daejeon 34055, Republic of Korea}\par
\vspace{0.55em}
\noindent\textbf{Junhan Kim} \hfill {\scriptsize ORCID: \texttt{0000-0002-4274-9373}}\par
{\footnotesize Department of Physics, Korea Advanced Institute of Science and Technology (KAIST), 291 Daehak-ro, Yuseong-gu, Daejeon 34141, Republic of Korea}\par
\vspace{0.55em}
\noindent\textbf{Motoki Kino} \hfill {\scriptsize ORCID: \texttt{0000-0002-2709-7338}}\par
{\footnotesize National Astronomical Observatory of Japan, 2-21-1 Osawa, Mitaka, Tokyo 181-8588, Japan}\par
{\footnotesize Kogakuin University of Technology \& Engineering, Academic Support Center, 2665-1 Nakano, Hachioji, Tokyo 192-0015, Japan}\par
\vspace{0.55em}
\noindent\textbf{Jakob Knollmüller} \hfill {\scriptsize ORCID: \texttt{0000-0002-9906-0040}}\par
{\footnotesize Department of Astrophysics, Institute for Mathematics, Astrophysics and Particle Physics (IMAPP), Radboud University, P.O. Box 9010, 6500 GL Nijmegen, The Netherlands}\par
{\footnotesize Max-Planck-Institut für Astrophysik, Karl-Schwarzschild-Str. 1, 85748 Garching, Germany}\par
{\footnotesize Technische Universität München, School of Computation, Information and Technology, Arcisstraße 21, 80333 Munich, Germany}\par
\vspace{0.55em}
\noindent\textbf{Jun Yi Koay} \hfill {\scriptsize ORCID: \texttt{0000-0002-7029-6658}}\par
{\footnotesize Graduate School of Science and Technology, Niigata University, 8050 Ikarashi 2-no-cho, Nishi-ku, Niigata 950-2181, Japan}\par
{\footnotesize Institute of Astronomy and Astrophysics, Academia Sinica, 11F of Astronomy-Mathematics Building, AS/NTU No. 1, Sec. 4, Roosevelt Rd., Taipei 106216, Taiwan, R.O.C.}\par
\vspace{0.55em}
\noindent\textbf{Prashant Kocherlakota} \hfill {\scriptsize ORCID: \texttt{0000-0001-7386-7439}}\par
{\footnotesize Black Hole Initiative at Harvard University, 20 Garden Street, Cambridge, MA 02138, USA}\par
{\footnotesize Center for Astrophysics $|$ Harvard \& Smithsonian, 60 Garden Street, Cambridge, MA 02138, USA}\par
\vspace{0.55em}
\noindent\textbf{Yutaro Kofuji}\par
{\footnotesize Instituto de Astrofísica de Andalucía-CSIC, Glorieta de la Astronomía s/n, E-18008 Granada, Spain}\par
{\footnotesize Mizusawa VLBI Observatory, National Astronomical Observatory of Japan, 2-12 Hoshigaoka, Mizusawa, Oshu, Iwate 023-0861, Japan}\par
\vspace{0.55em}
\noindent\textbf{Patrick M. Koch} \hfill {\scriptsize ORCID: \texttt{0000-0003-2777-5861}}\par
{\footnotesize Institute of Astronomy and Astrophysics, Academia Sinica, 11F of Astronomy-Mathematics Building, AS/NTU No. 1, Sec. 4, Roosevelt Rd., Taipei 106216, Taiwan, R.O.C.}\par
\vspace{0.55em}
\noindent\textbf{Shoko Koyama} \hfill {\scriptsize ORCID: \texttt{0000-0002-3723-3372}}\par
{\footnotesize Graduate School of Science and Technology, Niigata University, 8050 Ikarashi 2-no-cho, Nishi-ku, Niigata 950-2181, Japan}\par
{\footnotesize Institute of Astronomy and Astrophysics, Academia Sinica, 11F of Astronomy-Mathematics Building, AS/NTU No. 1, Sec. 4, Roosevelt Rd., Taipei 106216, Taiwan, R.O.C.}\par
\vspace{0.55em}
\noindent\textbf{Carsten Kramer} \hfill {\scriptsize ORCID: \texttt{0000-0002-4908-4925}}\par
{\footnotesize Institut de Radioastronomie Millimétrique (IRAM), 300 rue de la Piscine, F-38400 Saint-Martin-d'Hères, France}\par
\vspace{0.55em}
\noindent\textbf{Joana A. Kramer} \hfill {\scriptsize ORCID: \texttt{0009-0003-3011-0454}}\par
{\footnotesize Anton Pannekoek Institute for Astronomy, University of Amsterdam, Science Park 904, 1098 XH, Amsterdam, The Netherlands}\par
{\footnotesize Max-Planck-Institut für Radioastronomie, Auf dem Hügel 69, D-53121 Bonn, Germany}\par
\vspace{0.55em}
\noindent\textbf{Michael Kramer} \hfill {\scriptsize ORCID: \texttt{0000-0002-4175-2271}}\par
{\footnotesize Max-Planck-Institut für Radioastronomie, Auf dem Hügel 69, D-53121 Bonn, Germany}\par
\vspace{0.55em}
\noindent\textbf{Thomas P. Krichbaum} \hfill {\scriptsize ORCID: \texttt{0000-0002-4892-9586}}\par
{\footnotesize Max-Planck-Institut für Radioastronomie, Auf dem Hügel 69, D-53121 Bonn, Germany}\par
\vspace{0.55em}
\noindent\textbf{Cheng-Yu Kuo} \hfill {\scriptsize ORCID: \texttt{0000-0001-6211-5581}}\par
{\footnotesize Physics Department, National Sun Yat-Sen University, No. 70, Lien-Hai Road, Kaosiung City 80424, Taiwan, R.O.C.}\par
{\footnotesize Institute of Astronomy and Astrophysics, Academia Sinica, 11F of Astronomy-Mathematics Building, AS/NTU No. 1, Sec. 4, Roosevelt Rd., Taipei 106216, Taiwan, R.O.C.}\par
\vspace{0.55em}
\noindent\textbf{Noemi La Bella} \hfill {\scriptsize ORCID: \texttt{0000-0002-8116-9427}}\par
{\footnotesize Department of Astrophysics, Institute for Mathematics, Astrophysics and Particle Physics (IMAPP), Radboud University, P.O. Box 9010, 6500 GL Nijmegen, The Netherlands}\par
\vspace{0.55em}
\noindent\textbf{Deokhyeong Lee} \hfill {\scriptsize ORCID: \texttt{0009-0003-2122-9437}}\par
{\footnotesize Department of Astronomy, Kyungpook National University, 80 Daehak-ro, Buk-gu, Daegu 41566, Republic of Korea}\par
\vspace{0.55em}
\noindent\textbf{Sang-Sung Lee} \hfill {\scriptsize ORCID: \texttt{0000-0002-6269-594X}}\par
{\footnotesize Korea Astronomy and Space Science Institute, Daedeok-daero 776, Yuseong-gu, Daejeon 34055, Republic of Korea}\par
\vspace{0.55em}
\noindent\textbf{Aviad Levis} \hfill {\scriptsize ORCID: \texttt{0000-0001-7307-632X}}\par
{\footnotesize Department of Computer Science, University of Toronto, 40 St. George St., Toronto, ON, M5S 2E4, Canada}\par
{\footnotesize David A. Dunlap Department of Astronomy \& Astrophysics, University of Toronto, 50 St. George St, M5S 3H4, ON, Canada}\par
{\footnotesize Dunlap Institute for Astronomy and Astrophysics, University of Toronto, 50 St. George Street, Toronto, ON M5S 3H4, Canada}\par
\vspace{0.55em}
\noindent\textbf{Shaoliang Li} \hfill {\scriptsize ORCID: \texttt{0009-0005-0338-9490}}\par
{\footnotesize East Asian Observatory, 660 N. A'ohoku Place, Hilo, HI 96720, USA}\par
{\footnotesize James Clerk Maxwell Telescope (JCMT), 660 N. A'ohoku Place, Hilo, HI 96720, USA}\par
\vspace{0.55em}
\noindent\textbf{Zhiyuan Li (\cntext{李志远})} \hfill {\scriptsize ORCID: \texttt{0000-0003-0355-6437}}\par
{\footnotesize School of Astronomy and Space Science, Nanjing University, Nanjing 210023, People's Republic of China}\par
{\footnotesize Key Laboratory of Modern Astronomy and Astrophysics, Nanjing University, Nanjing 210023, People's Republic of China}\par
\vspace{0.55em}
\noindent\textbf{Rocco Lico} \hfill {\scriptsize ORCID: \texttt{0000-0001-7361-2460}}\par
{\footnotesize INAF-Istituto di Radioastronomia, Via P. Gobetti 101, I-40129 Bologna, Italy}\par
{\footnotesize Instituto de Astrofísica de Andalucía-CSIC, Glorieta de la Astronomía s/n, E-18008 Granada, Spain}\par
\vspace{0.55em}
\noindent\textbf{Greg Lindahl} \hfill {\scriptsize ORCID: \texttt{0000-0002-6100-4772}}\par
{\footnotesize Common Crawl Foundation, 9663 Santa Monica Blvd. 425, Beverly Hills, CA 90210 USA}\par
\vspace{0.55em}
\noindent\textbf{Michael Lindqvist} \hfill {\scriptsize ORCID: \texttt{0000-0002-3669-0715}}\par
{\footnotesize Department of Physics and Astronomy, Chalmers University of Technology, Onsala Space Observatory, SE-439 92 Onsala, Sweden}\par
\vspace{0.55em}
\noindent\textbf{Mikhail Lisakov} \hfill {\scriptsize ORCID: \texttt{0000-0001-6088-3819}}\par
{\footnotesize Instituto de Física, Pontificia Universidad Católica de Valparaíso, Casilla 4059, Valparaíso, Chile}\par
\vspace{0.55em}
\noindent\textbf{Jun Liu (\cntext{刘俊})} \hfill {\scriptsize ORCID: \texttt{0000-0002-7615-7499}}\par
{\footnotesize Max-Planck-Institut für Radioastronomie, Auf dem Hügel 69, D-53121 Bonn, Germany}\par
\vspace{0.55em}
\noindent\textbf{Kuo Liu} \hfill {\scriptsize ORCID: \texttt{0000-0002-2953-7376}}\par
{\footnotesize Key Laboratory of Radio Astronomy and Technology, Shanghai Astronomical Observatory, CAS, 80 Nandan Road, Shanghai 200030, People’s Republic of China}\par
\vspace{0.55em}
\noindent\textbf{Elisabetta Liuzzo} \hfill {\scriptsize ORCID: \texttt{0000-0003-0995-5201}}\par
{\footnotesize INAF-Istituto di Radioastronomia \& Italian ALMA Regional Centre, Via P. Gobetti 101, I-40129 Bologna, Italy}\par
\vspace{0.55em}
\noindent\textbf{Wen-Ping Lo} \hfill {\scriptsize ORCID: \texttt{0000-0003-1869-2503}}\par
{\footnotesize Institute of Astronomy and Astrophysics, Academia Sinica, 11F of Astronomy-Mathematics Building, AS/NTU No. 1, Sec. 4, Roosevelt Rd., Taipei 106216, Taiwan, R.O.C.}\par
{\footnotesize Department of Physics, National Taiwan University, No. 1, Sec. 4, Roosevelt Rd., Taipei 106216, Taiwan, R.O.C}\par
\vspace{0.55em}
\noindent\textbf{Andrei P. Lobanov} \hfill {\scriptsize ORCID: \texttt{0000-0003-1622-1484}}\par
{\footnotesize Max-Planck-Institut für Radioastronomie, Auf dem Hügel 69, D-53121 Bonn, Germany}\par
\vspace{0.55em}
\noindent\textbf{Laurent Loinard} \hfill {\scriptsize ORCID: \texttt{0000-0002-5635-3345}}\par
{\footnotesize Instituto de Radioastronomía y Astrofísica, Universidad Nacional Autónoma de México, Morelia 58089, México}\par
{\footnotesize Black Hole Initiative at Harvard University, 20 Garden Street, Cambridge, MA 02138, USA}\par
\vspace{0.55em}
\noindent\textbf{Colin J. Lonsdale} \hfill {\scriptsize ORCID: \texttt{0000-0003-4062-4654}}\par
{\footnotesize Massachusetts Institute of Technology Haystack Observatory, 99 Millstone Road, Westford, MA 01886, USA}\par
\vspace{0.55em}
\noindent\textbf{Amy E. Lowitz} \hfill {\scriptsize ORCID: \texttt{0000-0002-4747-4276}}\par
{\footnotesize Steward Observatory and Department of Astronomy, University of Arizona, 933 N. Cherry Ave., Tucson, AZ 85721, USA}\par
\vspace{0.55em}
\noindent\textbf{Ru-Sen Lu (\cntext{路如森})} \hfill {\scriptsize ORCID: \texttt{0000-0002-7692-7967}}\par
{\footnotesize Shanghai Astronomical Observatory, Chinese Academy of Sciences, 80 Nandan Road, Shanghai 200030, People's Republic of China}\par
{\footnotesize Key Laboratory of Radio Astronomy and Technology, Chinese Academy of Sciences, A20 Datun Road, Chaoyang District, Beijing, 100101, People’s Republic of China}\par
{\footnotesize Max-Planck-Institut für Radioastronomie, Auf dem Hügel 69, D-53121 Bonn, Germany}\par
\vspace{0.55em}
\noindent\textbf{Nicholas R. MacDonald} \hfill {\scriptsize ORCID: \texttt{0000-0002-6684-8691}}\par
{\footnotesize Department of Physics and Astronomy, University of Mississippi, Mississippi 38677, USA}\par
{\footnotesize Max-Planck-Institut für Radioastronomie, Auf dem Hügel 69, D-53121 Bonn, Germany}\par
\vspace{0.55em}
\noindent\textbf{Jirong Mao (\cntext{毛基荣})} \hfill {\scriptsize ORCID: \texttt{0000-0002-7077-7195}}\par
{\footnotesize Yunnan Observatories, Chinese Academy of Sciences, 650011 Kunming, Yunnan Province, People's Republic of China}\par
{\footnotesize Center for Astronomical Mega-Science, Chinese Academy of Sciences, 20A Datun Road, Chaoyang District, Beijing, 100012, People's Republic of China}\par
{\footnotesize Key Laboratory for the Structure and Evolution of Celestial Objects, Chinese Academy of Sciences, 650011 Kunming, People's Republic of China}\par
\vspace{0.55em}
\noindent\textbf{Nicola Marchili} \hfill {\scriptsize ORCID: \texttt{0000-0002-5523-7588}}\par
{\footnotesize INAF-Istituto di Radioastronomia \& Italian ALMA Regional Centre, Via P. Gobetti 101, I-40129 Bologna, Italy}\par
{\footnotesize Max-Planck-Institut für Radioastronomie, Auf dem Hügel 69, D-53121 Bonn, Germany}\par
\vspace{0.55em}
\noindent\textbf{Sera Markoff} \hfill {\scriptsize ORCID: \texttt{0000-0001-9564-0876}}\par
{\footnotesize Anton Pannekoek Institute for Astronomy, University of Amsterdam, Science Park 904, 1098 XH, Amsterdam, The Netherlands}\par
{\footnotesize Gravitation and Astroparticle Physics Amsterdam (GRAPPA) Institute, University of Amsterdam, Science Park 904, 1098 XH Amsterdam, The Netherlands}\par
{\footnotesize Institute of Astronomy, University of Cambridge, Madingley Road, Cambridge CB3 0HA, United Kingdom}\par
\vspace{0.55em}
\noindent\textbf{Daniel P. Marrone} \hfill {\scriptsize ORCID: \texttt{0000-0002-2367-1080}}\par
{\footnotesize Steward Observatory and Department of Astronomy, University of Arizona, 933 N. Cherry Ave., Tucson, AZ 85721, USA}\par
\vspace{0.55em}
\noindent\textbf{Alan P. Marscher} \hfill {\scriptsize ORCID: \texttt{0000-0001-7396-3332}}\par
{\footnotesize Institute for Astrophysical Research, Boston University, 725 Commonwealth Ave., Boston, MA 02215, USA}\par
\vspace{0.55em}
\noindent\textbf{Iván Martí-Vidal} \hfill {\scriptsize ORCID: \texttt{0000-0003-3708-9611}}\par
{\footnotesize Departament d'Astronomia i Astrofísica, Universitat de València, C. Dr. Moliner 50, E-46100 Burjassot, València, Spain}\par
{\footnotesize Observatori Astronòmic, Universitat de València, C. Catedrático José Beltrán 2, E-46980 Paterna, València, Spain}\par
\vspace{0.55em}
\noindent\textbf{Satoki Matsushita} \hfill {\scriptsize ORCID: \texttt{0000-0002-2127-7880}}\par
{\footnotesize Institute of Astronomy and Astrophysics, Academia Sinica, 11F of Astronomy-Mathematics Building, AS/NTU No. 1, Sec. 4, Roosevelt Rd., Taipei 106216, Taiwan, R.O.C.}\par
\vspace{0.55em}
\noindent\textbf{Lynn D. Matthews} \hfill {\scriptsize ORCID: \texttt{0000-0002-3728-8082}}\par
{\footnotesize Massachusetts Institute of Technology Haystack Observatory, 99 Millstone Road, Westford, MA 01886, USA}\par
\vspace{0.55em}
\noindent\textbf{Lia Medeiros} \hfill {\scriptsize ORCID: \texttt{0000-0003-2342-6728}}\par
{\footnotesize Center for Gravitation, Cosmology and Astrophysics, Department of Physics, University of Wisconsin–Milwaukee, P.O. Box 413, Milwaukee, WI 53201, USA}\par
\vspace{0.55em}
\noindent\textbf{Karl M. Menten} \hfill {\scriptsize ORCID: \texttt{0000-0001-6459-0669}}\par
{\footnotesize Max-Planck-Institut für Radioastronomie, Auf dem Hügel 69, D-53121 Bonn, Germany}\par
{\footnotesize Deceased}\par
\vspace{0.55em}
\noindent\textbf{Hugo Messias} \hfill {\scriptsize ORCID: \texttt{0000-0002-2985-7994}}\par
{\footnotesize Joint ALMA Observatory, Alonso de C\'ordova 3107, Vitacura 763-0355, Santiago, Chile}\par
{\footnotesize European Southern Observatory, Alonso de C\'ordova 3107, Vitacura, Casilla 19001, Santiago, Chile}\par
\vspace{0.55em}
\noindent\textbf{Izumi Mizuno} \hfill {\scriptsize ORCID: \texttt{0000-0002-7210-6264}}\par
{\footnotesize East Asian Observatory, 660 N. A'ohoku Place, Hilo, HI 96720, USA}\par
{\footnotesize James Clerk Maxwell Telescope (JCMT), 660 N. A'ohoku Place, Hilo, HI 96720, USA}\par
\vspace{0.55em}
\noindent\textbf{Yosuke Mizuno} \hfill {\scriptsize ORCID: \texttt{0000-0002-8131-6730}}\par
{\footnotesize Tsung-Dao Lee Institute, Shanghai Jiao Tong University, Shengrong Road 520, Shanghai, 201210, People’s Republic of China}\par
{\footnotesize School of Physics and Astronomy, Shanghai Jiao Tong University, 800 Dongchuan Road, Shanghai, 200240, People’s Republic of China}\par
{\footnotesize Institut für Theoretische Physik, Goethe-Universität Frankfurt, Max-von-Laue-Straße 1, D-60438 Frankfurt am Main, Germany}\par
\vspace{0.55em}
\noindent\textbf{Joshua Montgomery} \hfill {\scriptsize ORCID: \texttt{0000-0003-0345-8386}}\par
{\footnotesize Trottier Space Institute at McGill, 3550 rue University, Montréal, QC H3A 2A7, Canada}\par
{\footnotesize Department of Astronomy and Astrophysics, University of Chicago, 5640 South Ellis Avenue, Chicago, IL 60637, USA}\par
\vspace{0.55em}
\noindent\textbf{Kotaro Moriyama} \hfill {\scriptsize ORCID: \texttt{0000-0003-1364-3761}}\par
{\footnotesize Instituto de Astrofísica de Andalucía-CSIC, Glorieta de la Astronomía s/n, E-18008 Granada, Spain}\par
{\footnotesize Mizusawa VLBI Observatory, National Astronomical Observatory of Japan, 2-12 Hoshigaoka, Mizusawa, Oshu, Iwate 023-0861, Japan}\par
\vspace{0.55em}
\noindent\textbf{Monika Moscibrodzka} \hfill {\scriptsize ORCID: \texttt{0000-0002-4661-6332}}\par
{\footnotesize Department of Astrophysics, Institute for Mathematics, Astrophysics and Particle Physics (IMAPP), Radboud University, P.O. Box 9010, 6500 GL Nijmegen, The Netherlands}\par
\vspace{0.55em}
\noindent\textbf{Wanga Mulaudzi} \hfill {\scriptsize ORCID: \texttt{0000-0003-4514-625X}}\par
{\footnotesize Anton Pannekoek Institute for Astronomy, University of Amsterdam, Science Park 904, 1098 XH, Amsterdam, The Netherlands}\par
\vspace{0.55em}
\noindent\textbf{Cornelia Müller} \hfill {\scriptsize ORCID: \texttt{0000-0002-2739-2994}}\par
{\footnotesize Max-Planck-Institut für Radioastronomie, Auf dem Hügel 69, D-53121 Bonn, Germany}\par
{\footnotesize Department of Astrophysics, Institute for Mathematics, Astrophysics and Particle Physics (IMAPP), Radboud University, P.O. Box 9010, 6500 GL Nijmegen, The Netherlands}\par
\vspace{0.55em}
\noindent\textbf{Hendrik Müller} \hfill {\scriptsize ORCID: \texttt{0000-0002-9250-0197}}\par
{\footnotesize National Radio Astronomy Observatory, 1011 Lopezville Road, 87801 Socorro, NM, USA}\par
{\footnotesize Max-Planck-Institut für Radioastronomie, Auf dem Hügel 69, D-53121 Bonn, Germany}\par
\vspace{0.55em}
\noindent\textbf{Alejandro Mus} \hfill {\scriptsize ORCID: \texttt{0000-0003-0329-6874}}\par
{\footnotesize Dipartimento di Fisica, Università degli Studi di Cagliari, SP Monserrato-Sestu km 0.7, I-09042 Monserrato (CA), Italy}\par
{\footnotesize INAF-Istituto di Radioastronomia, Via P. Gobetti 101, I-40129 Bologna, Italy}\par
{\footnotesize Soft Computing, Image Processing and Aggregation Research Group (SCOPIA) \& Modelling and Imaging Radio Astronomical Data (MIRADA), Dept. of Mathematics and Computer Science, University of the Balearic Islands, Ctra. Valldemossa, Km 7.5, Palma 07122, Spain}\par
{\footnotesize Artificial Intelligence Research Institute of the Balearic Islands (IAIB), Palma 07122, Spain}\par
{\footnotesize Health Research Institute of the Balearic Islands (IdISBa), 07010 Palma de Mallorca, Spain}\par
\vspace{0.55em}
\noindent\textbf{Gibwa Musoke} \hfill {\scriptsize ORCID: \texttt{0000-0003-1984-189X}}\par
{\footnotesize Anton Pannekoek Institute for Astronomy, University of Amsterdam, Science Park 904, 1098 XH, Amsterdam, The Netherlands}\par
{\footnotesize Department of Astrophysics, Institute for Mathematics, Astrophysics and Particle Physics (IMAPP), Radboud University, P.O. Box 9010, 6500 GL Nijmegen, The Netherlands}\par
\vspace{0.55em}
\noindent\textbf{Ioannis Myserlis} \hfill {\scriptsize ORCID: \texttt{0000-0003-3025-9497}}\par
{\footnotesize Institut de Radioastronomie Millimétrique (IRAM), Avenida Divina Pastora 7, Local 20, E-18012, Granada, Spain}\par
\vspace{0.55em}
\noindent\textbf{Hiroshi Nagai} \hfill {\scriptsize ORCID: \texttt{0000-0003-0292-3645}}\par
{\footnotesize National Astronomical Observatory of Japan, 2-21-1 Osawa, Mitaka, Tokyo 181-8588, Japan}\par
{\footnotesize Department of Astronomical Science, The Graduate University for Advanced Studies (SOKENDAI), 2-21-1 Osawa, Mitaka, Tokyo 181-8588, Japan}\par
\vspace{0.55em}
\noindent\textbf{Neil M. Nagar} \hfill {\scriptsize ORCID: \texttt{0000-0001-6920-662X}}\par
{\footnotesize Astronomy Department, Universidad de Concepción, Casilla 160-C, Concepción, Chile}\par
\vspace{0.55em}
\noindent\textbf{Dhanya G. Nair} \hfill {\scriptsize ORCID: \texttt{0000-0001-5357-7805}}\par
{\footnotesize Astronomy Department, Universidad de Concepción, Casilla 160-C, Concepción, Chile}\par
{\footnotesize Max-Planck-Institut für Radioastronomie, Auf dem Hügel 69, D-53121 Bonn, Germany}\par
\vspace{0.55em}
\noindent\textbf{Masanori Nakamura} \hfill {\scriptsize ORCID: \texttt{0000-0001-6081-2420}}\par
{\footnotesize National Institute of Technology, Hachinohe College, 16-1 Uwanotai, Tamonoki, Hachinohe City, Aomori 039-1192, Japan}\par
{\footnotesize Institute of Astronomy and Astrophysics, Academia Sinica, 11F of Astronomy-Mathematics Building, AS/NTU No. 1, Sec. 4, Roosevelt Rd., Taipei 106216, Taiwan, R.O.C.}\par
\vspace{0.55em}
\noindent\textbf{Gopal Narayanan} \hfill {\scriptsize ORCID: \texttt{0000-0002-4723-6569}}\par
{\footnotesize Department of Astronomy, University of Massachusetts, Amherst, MA 01003, USA}\par
\vspace{0.55em}
\noindent\textbf{Iniyan Natarajan} \hfill {\scriptsize ORCID: \texttt{0000-0001-8242-4373}}\par
{\footnotesize SKA Observatory, Jodrell Bank, Lower Withington, Macclesfield, SK11 9FT, UK}\par
{\footnotesize Center for Astrophysics $|$ Harvard \& Smithsonian, 60 Garden Street, Cambridge, MA 02138, USA}\par
{\footnotesize Black Hole Initiative at Harvard University, 20 Garden Street, Cambridge, MA 02138, USA}\par
\vspace{0.55em}
\noindent\textbf{Antonios Nathanail} \hfill {\scriptsize ORCID: \texttt{0000-0002-1655-9912}}\par
{\footnotesize Research Center for Astronomy, Academy of Athens, Soranou Efessiou 4, 115 27 Athens, Greece}\par
\vspace{0.55em}
\noindent\textbf{Santiago Navarro Fuentes}\par
{\footnotesize Institut de Radioastronomie Millimétrique (IRAM), Avenida Divina Pastora 7, Local 20, E-18012, Granada, Spain}\par
\vspace{0.55em}
\noindent\textbf{Joey Neilsen} \hfill {\scriptsize ORCID: \texttt{0000-0002-8247-786X}}\par
{\footnotesize Department of Physics, Villanova University, 800 Lancaster Avenue, Villanova, PA 19085, USA}\par
\vspace{0.55em}
\noindent\textbf{Chunchong Ni} \hfill {\scriptsize ORCID: \texttt{0000-0003-1361-5699}}\par
{\footnotesize Department of Physics and Astronomy, University of Waterloo, 200 University Avenue West, Waterloo, ON N2L 3G1, Canada}\par
{\footnotesize Waterloo Centre for Astrophysics, University of Waterloo, Waterloo, ON N2L 3G1, Canada}\par
{\footnotesize Perimeter Institute for Theoretical Physics, 31 Caroline Street North, Waterloo, ON N2L 2Y5, Canada}\par
\vspace{0.55em}
\noindent\textbf{Andy Nilipour} \hfill {\scriptsize ORCID: \texttt{0000-0002-5956-5167}}\par
{\footnotesize Cavendish Astrophysics, University of Cambridge, Madingley Road, Cambridge CB3 0HA, UK}\par
{\footnotesize Kavli Institute for Cosmology, University of Cambridge, Madingley Road, Cambridge CB3 0HA, UK}\par
\vspace{0.55em}
\noindent\textbf{Michael A. Nowak} \hfill {\scriptsize ORCID: \texttt{0000-0001-6923-1315}}\par
{\footnotesize Physics Department, Washington University, CB 1105, St. Louis, MO 63130, USA}\par
\vspace{0.55em}
\noindent\textbf{Hiroki Okino} \hfill {\scriptsize ORCID: \texttt{0000-0003-3779-2016}}\par
{\footnotesize Mizusawa VLBI Observatory, National Astronomical Observatory of Japan, 2-12 Hoshigaoka, Mizusawa, Oshu, Iwate 023-0861, Japan}\par
{\footnotesize Department of Astronomy, Graduate School of Science, The University of Tokyo, 7-3-1 Hongo, Bunkyo-ku, Tokyo 113-0033, Japan}\par
\vspace{0.55em}
\noindent\textbf{Héctor Raúl Olivares Sánchez} \hfill {\scriptsize ORCID: \texttt{0000-0001-6833-7580}}\par
{\footnotesize Departamento de Matemática da Universidade de Aveiro and Centre for Research and Development in Mathematics and Applications (CIDMA), Campus de Santiago, 3810-193 Aveiro, Portugal}\par
\vspace{0.55em}
\noindent\textbf{Feryal Özel} \hfill {\scriptsize ORCID: \texttt{0000-0003-4413-1523}}\par
{\footnotesize School of Physics, Georgia Institute of Technology, 837 State St NW, Atlanta, GA 30332, USA}\par
\vspace{0.55em}
\noindent\textbf{Daniel C. M. Palumbo} \hfill {\scriptsize ORCID: \texttt{0000-0002-7179-3816}}\par
{\footnotesize Black Hole Initiative at Harvard University, 20 Garden Street, Cambridge, MA 02138, USA}\par
{\footnotesize Center for Astrophysics $|$ Harvard \& Smithsonian, 60 Garden Street, Cambridge, MA 02138, USA}\par
\vspace{0.55em}
\noindent\textbf{Georgios Filippos Paraschos} \hfill {\scriptsize ORCID: \texttt{0000-0001-6757-3098}}\par
{\footnotesize Finnish Centre for Astronomy with ESO, University of Turku, FI-20014 Turun Yliopisto, Finland}\par
{\footnotesize Aalto University Metsähovi Radio Observatory, Metsähovintie 114, FI-02540 Kylmälä, Finland}\par
{\footnotesize Max-Planck-Institut für Radioastronomie, Auf dem Hügel 69, D-53121 Bonn, Germany}\par
\vspace{0.55em}
\noindent\textbf{Jongho Park} \hfill {\scriptsize ORCID: \texttt{0000-0001-6558-9053}}\par
{\footnotesize School of Space Research, Kyung Hee University, 1732, Deogyeong-daero, Giheung-gu, Yongin-si, Gyeonggi-do 17104, Republic of Korea}\par
{\footnotesize G-LAMP NEXUS Institute, Kyung Hee University, Yongin, 17104, Republic of Korea}\par
{\footnotesize Institute of Astronomy and Astrophysics, Academia Sinica, 11F of Astronomy-Mathematics Building, AS/NTU No. 1, Sec. 4, Roosevelt Rd., Taipei 106216, Taiwan, R.O.C.}\par
\vspace{0.55em}
\noindent\textbf{Harriet Parsons} \hfill {\scriptsize ORCID: \texttt{0000-0002-6327-3423}}\par
{\footnotesize East Asian Observatory, 660 N. A'ohoku Place, Hilo, HI 96720, USA}\par
{\footnotesize James Clerk Maxwell Telescope (JCMT), 660 N. A'ohoku Place, Hilo, HI 96720, USA}\par
\vspace{0.55em}
\noindent\textbf{Nimesh Patel} \hfill {\scriptsize ORCID: \texttt{0000-0002-6021-9421}}\par
{\footnotesize Center for Astrophysics $|$ Harvard \& Smithsonian, 60 Garden Street, Cambridge, MA 02138, USA}\par
\vspace{0.55em}
\noindent\textbf{Ue-Li Pen} \hfill {\scriptsize ORCID: \texttt{0000-0003-2155-9578}}\par
{\footnotesize Institute of Astronomy and Astrophysics, Academia Sinica, 11F of Astronomy-Mathematics Building, AS/NTU No. 1, Sec. 4, Roosevelt Rd., Taipei 106216, Taiwan, R.O.C.}\par
{\footnotesize Perimeter Institute for Theoretical Physics, 31 Caroline Street North, Waterloo, ON N2L 2Y5, Canada}\par
{\footnotesize Canadian Institute for Theoretical Astrophysics, University of Toronto, 60 St. George Street, Toronto, ON M5S 3H8, Canada}\par
{\footnotesize Dunlap Institute for Astronomy and Astrophysics, University of Toronto, 50 St. George Street, Toronto, ON M5S 3H4, Canada}\par
{\footnotesize Canadian Institute for Advanced Research, 180 Dundas St West, Toronto, ON M5G 1Z8, Canada}\par
\vspace{0.55em}
\noindent\textbf{Dominic W. Pesce} \hfill {\scriptsize ORCID: \texttt{0000-0002-5278-9221}}\par
{\footnotesize Center for Astrophysics $|$ Harvard \& Smithsonian, 60 Garden Street, Cambridge, MA 02138, USA}\par
{\footnotesize Black Hole Initiative at Harvard University, 20 Garden Street, Cambridge, MA 02138, USA}\par
\vspace{0.55em}
\noindent\textbf{Vincent Piétu} \hfill {\scriptsize ORCID: \texttt{0009-0006-3497-397X}}\par
{\footnotesize Institut de Radioastronomie Millimétrique (IRAM), 300 rue de la Piscine, F-38400 Saint-Martin-d'Hères, France}\par
\vspace{0.55em}
\noindent\textbf{Alexander Plavin} \hfill {\scriptsize ORCID: \texttt{0000-0003-2914-8554}}\par
{\footnotesize Black Hole Initiative at Harvard University, 20 Garden Street, Cambridge, MA 02138, USA}\par
{\footnotesize Center for Astrophysics $|$ Harvard \& Smithsonian, 60 Garden Street, Cambridge, MA 02138, USA}\par
{\footnotesize Max-Planck-Institut für Radioastronomie, Auf dem Hügel 69, D-53121 Bonn, Germany}\par
\vspace{0.55em}
\noindent\textbf{Aleksandar PopStefanija}\par
{\footnotesize Department of Astronomy, University of Massachusetts, Amherst, MA 01003, USA}\par
\vspace{0.55em}
\noindent\textbf{Oliver Porth} \hfill {\scriptsize ORCID: \texttt{0000-0002-4584-2557}}\par
{\footnotesize Anton Pannekoek Institute for Astronomy, University of Amsterdam, Science Park 904, 1098 XH, Amsterdam, The Netherlands}\par
{\footnotesize Institut für Theoretische Physik, Goethe-Universität Frankfurt, Max-von-Laue-Straße 1, D-60438 Frankfurt am Main, Germany}\par
\vspace{0.55em}
\noindent\textbf{Cora Prather} \hfill {\scriptsize ORCID: \texttt{0000-0002-0393-7734}}\par
{\footnotesize Black Hole Initiative at Harvard University, 20 Garden Street, Cambridge, MA 02138, USA}\par
\vspace{0.55em}
\noindent\textbf{Giacomo Principe} \hfill {\scriptsize ORCID: \texttt{0000-0003-0406-7387}}\par
{\footnotesize Dipartimento di Fisica, Università di Trieste, I-34127 Trieste, Italy}\par
{\footnotesize INFN Sez. di Trieste, I-34127 Trieste, Italy}\par
{\footnotesize INAF-Istituto di Radioastronomia, Via P. Gobetti 101, I-40129 Bologna, Italy}\par
\vspace{0.55em}
\noindent\textbf{Dimitrios Psaltis} \hfill {\scriptsize ORCID: \texttt{0000-0003-1035-3240}}\par
{\footnotesize School of Physics, Georgia Institute of Technology, 837 State St NW, Atlanta, GA 30332, USA}\par
\vspace{0.55em}
\noindent\textbf{Hung-Yi Pu} \hfill {\scriptsize ORCID: \texttt{0000-0001-9270-8812}}\par
{\footnotesize Department of Physics, National Taiwan Normal University, No. 88, Sec. 4, Tingzhou Rd., Taipei 116, Taiwan, R.O.C.}\par
{\footnotesize Center of Astronomy and Gravitation, National Taiwan Normal University, No. 88, Sec. 4, Tingzhou Road, Taipei 116, Taiwan, R.O.C.}\par
{\footnotesize Institute of Astronomy and Astrophysics, Academia Sinica, 11F of Astronomy-Mathematics Building, AS/NTU No. 1, Sec. 4, Roosevelt Rd., Taipei 106216, Taiwan, R.O.C.}\par
\vspace{0.55em}
\noindent\textbf{Alexandra Rahlin} \hfill {\scriptsize ORCID: \texttt{0000-0003-3953-1776}}\par
{\footnotesize Department of Astronomy and Astrophysics, University of Chicago, 5640 South Ellis Avenue, Chicago, IL 60637, USA}\par
\vspace{0.55em}
\noindent\textbf{Venkatessh Ramakrishnan} \hfill {\scriptsize ORCID: \texttt{0000-0002-9248-086X}}\par
{\footnotesize Signal Processing Research Centre, Tampere University, FI-33720 Tampere, Finland}\par
{\footnotesize Finnish Centre for Astronomy with ESO, University of Turku, FI-20014 Turun Yliopisto, Finland}\par
{\footnotesize Aalto University Metsähovi Radio Observatory, Metsähovintie 114, FI-02540 Kylmälä, Finland}\par
\vspace{0.55em}
\noindent\textbf{Ramprasad Rao} \hfill {\scriptsize ORCID: \texttt{0000-0002-1407-7944}}\par
{\footnotesize Center for Astrophysics $|$ Harvard \& Smithsonian, 60 Garden Street, Cambridge, MA 02138, USA}\par
\vspace{0.55em}
\noindent\textbf{Mark G. Rawlings} \hfill {\scriptsize ORCID: \texttt{0000-0002-6529-202X}}\par
{\footnotesize National Radio Astronomy Observatory, 520 Edgemont Road, Charlottesville, VA 22903, USA}\par
{\footnotesize East Asian Observatory, 660 N. A'ohoku Place, Hilo, HI 96720, USA}\par
{\footnotesize James Clerk Maxwell Telescope (JCMT), 660 N. A'ohoku Place, Hilo, HI 96720, USA}\par
\vspace{0.55em}
\noindent\textbf{Luciano Rezzolla} \hfill {\scriptsize ORCID: \texttt{0000-0002-1330-7103}}\par
{\footnotesize Institut für Theoretische Physik, Goethe-Universität Frankfurt, Max-von-Laue-Straße 1, D-60438 Frankfurt am Main, Germany}\par
{\footnotesize Department of Mathematics, New Uzbekistan University, Tashkent 100007, Uzbekistan}\par
{\footnotesize School of Mathematics, Trinity College, Dublin 2, Ireland}\par
\vspace{0.55em}
\noindent\textbf{Angelo Ricarte} \hfill {\scriptsize ORCID: \texttt{0000-0001-5287-0452}}\par
{\footnotesize Black Hole Initiative at Harvard University, 20 Garden Street, Cambridge, MA 02138, USA}\par
{\footnotesize Center for Astrophysics $|$ Harvard \& Smithsonian, 60 Garden Street, Cambridge, MA 02138, USA}\par
\vspace{0.55em}
\noindent\textbf{Luca Ricci} \hfill {\scriptsize ORCID: \texttt{0000-0002-4175-3194}}\par
{\footnotesize Julius-Maximilians-Universität Würzburg, Fakultät für Physik und Astronomie, Institut für Theoretische Physik und Astrophysik, Lehrstuhl für Astronomie, Emil-Fischer-Str. 31, D-97074 Würzburg, Germany}\par
\vspace{0.55em}
\noindent\textbf{Bart Ripperda} \hfill {\scriptsize ORCID: \texttt{0000-0002-7301-3908}}\par
{\footnotesize Canadian Institute for Theoretical Astrophysics, University of Toronto, 60 St. George Street, Toronto, ON M5S 3H8, Canada}\par
{\footnotesize Department of Physics, University of Toronto, 60 St. George Street, Toronto, ON M5S 1A7, Canada}\par
{\footnotesize Dunlap Institute for Astronomy and Astrophysics, University of Toronto, 50 St. George Street, Toronto, ON M5S 3H4, Canada}\par
{\footnotesize Perimeter Institute for Theoretical Physics, 31 Caroline Street North, Waterloo, ON N2L 2Y5, Canada}\par
\vspace{0.55em}
\noindent\textbf{Jan Röder} \hfill {\scriptsize ORCID: \texttt{0000-0002-2426-927X}}\par
{\footnotesize Instituto de Astrofísica de Andalucía-CSIC, Glorieta de la Astronomía s/n, E-18008 Granada, Spain}\par
\vspace{0.55em}
\noindent\textbf{Freek Roelofs} \hfill {\scriptsize ORCID: \texttt{0000-0001-5461-3687}}\par
{\footnotesize Department of Astrophysics, Institute for Mathematics, Astrophysics and Particle Physics (IMAPP), Radboud University, P.O. Box 9010, 6500 GL Nijmegen, The Netherlands}\par
\vspace{0.55em}
\noindent\textbf{Cristina Romero-Cañizales} \hfill {\scriptsize ORCID: \texttt{0000-0001-6301-9073}}\par
{\footnotesize Institute of Astronomy and Astrophysics, Academia Sinica, 11F of Astronomy-Mathematics Building, AS/NTU No. 1, Sec. 4, Roosevelt Rd., Taipei 106216, Taiwan, R.O.C.}\par
\vspace{0.55em}
\noindent\textbf{Eduardo Ros} \hfill {\scriptsize ORCID: \texttt{0000-0001-9503-4892}}\par
{\footnotesize Max-Planck-Institut für Radioastronomie, Auf dem Hügel 69, D-53121 Bonn, Germany}\par
\vspace{0.55em}
\noindent\textbf{Arash Roshanineshat} \hfill {\scriptsize ORCID: \texttt{0000-0002-8280-9238}}\par
{\footnotesize Steward Observatory and Department of Astronomy, University of Arizona, 933 N. Cherry Ave., Tucson, AZ 85721, USA}\par
\vspace{0.55em}
\noindent\textbf{Helge Rottmann}\par
{\footnotesize Max-Planck-Institut für Radioastronomie, Auf dem Hügel 69, D-53121 Bonn, Germany}\par
\vspace{0.55em}
\noindent\textbf{Alan L. Roy} \hfill {\scriptsize ORCID: \texttt{0000-0002-1931-0135}}\par
{\footnotesize Max-Planck-Institut für Radioastronomie, Auf dem Hügel 69, D-53121 Bonn, Germany}\par
\vspace{0.55em}
\noindent\textbf{Ignacio Ruiz} \hfill {\scriptsize ORCID: \texttt{0000-0002-0965-5463}}\par
{\footnotesize Institut de Radioastronomie Millimétrique (IRAM), Avenida Divina Pastora 7, Local 20, E-18012, Granada, Spain}\par
\vspace{0.55em}
\noindent\textbf{Chet Ruszczyk} \hfill {\scriptsize ORCID: \texttt{0000-0001-7278-9707}}\par
{\footnotesize Massachusetts Institute of Technology Haystack Observatory, 99 Millstone Road, Westford, MA 01886, USA}\par
\vspace{0.55em}
\noindent\textbf{Kazi L. J. Rygl} \hfill {\scriptsize ORCID: \texttt{0000-0003-4146-9043}}\par
{\footnotesize INAF-Istituto di Radioastronomia \& Italian ALMA Regional Centre, Via P. Gobetti 101, I-40129 Bologna, Italy}\par
\vspace{0.55em}
\noindent\textbf{León D. S. Salas} \hfill {\scriptsize ORCID: \texttt{0000-0003-1979-6363}}\par
{\footnotesize Anton Pannekoek Institute for Astronomy, University of Amsterdam, Science Park 904, 1098 XH, Amsterdam, The Netherlands}\par
\vspace{0.55em}
\noindent\textbf{Salvador Sánchez} \hfill {\scriptsize ORCID: \texttt{0000-0002-8042-5951}}\par
{\footnotesize Institut de Radioastronomie Millimétrique (IRAM), Avenida Divina Pastora 7, Local 20, E-18012, Granada, Spain}\par
\vspace{0.55em}
\noindent\textbf{David Sánchez-Argüelles} \hfill {\scriptsize ORCID: \texttt{0000-0002-7344-9920}}\par
{\footnotesize Instituto Nacional de Astrofísica, Óptica y Electrónica. Apartado Postal 51 y 216, 72000. Puebla Pue., México}\par
{\footnotesize Consejo Nacional de Humanidades, Ciencia y Tecnología, Av. Insurgentes Sur 1582, 03940, Ciudad de México, México}\par
\vspace{0.55em}
\noindent\textbf{Miguel Sánchez-Portal} \hfill {\scriptsize ORCID: \texttt{0000-0003-0981-9664}}\par
{\footnotesize Institut de Radioastronomie Millimétrique (IRAM), Avenida Divina Pastora 7, Local 20, E-18012, Granada, Spain}\par
\vspace{0.55em}
\noindent\textbf{Ali SaraerToosi} \hfill {\scriptsize ORCID: \texttt{0009-0003-4620-8448}}\par
{\footnotesize Department of Computer Science, University of Toronto, 40 St. George St., Toronto, ON, M5S 2E4, Canada}\par
\vspace{0.55em}
\noindent\textbf{Mahito Sasada} \hfill {\scriptsize ORCID: \texttt{0000-0001-5946-9960}}\par
{\footnotesize Department of Physics, Tokyo Institute of Technology, 2-12-1 Ookayama, Meguro-ku, Tokyo 152-8551, Japan}\par
{\footnotesize Mizusawa VLBI Observatory, National Astronomical Observatory of Japan, 2-12 Hoshigaoka, Mizusawa, Oshu, Iwate 023-0861, Japan}\par
{\footnotesize Hiroshima Astrophysical Science Center, Hiroshima University, 1-3-1 Kagamiyama, Higashi-Hiroshima, Hiroshima 739-8526, Japan}\par
\vspace{0.55em}
\noindent\textbf{Kaushik Satapathy} \hfill {\scriptsize ORCID: \texttt{0000-0003-0433-3585}}\par
{\footnotesize Steward Observatory and Department of Astronomy, University of Arizona, 933 N. Cherry Ave., Tucson, AZ 85721, USA}\par
\vspace{0.55em}
\noindent\textbf{Saurabh} \hfill {\scriptsize ORCID: \texttt{0000-0001-7156-4848}}\par
{\footnotesize Max-Planck-Institut für Radioastronomie, Auf dem Hügel 69, D-53121 Bonn, Germany}\par
\vspace{0.55em}
\noindent\textbf{Tuomas Savolainen} \hfill {\scriptsize ORCID: \texttt{0000-0001-6214-1085}}\par
{\footnotesize Aalto University Department of Electronics and Nanoengineering, PL 15500, FI-00076 Aalto, Finland}\par
{\footnotesize Aalto University Metsähovi Radio Observatory, Metsähovintie 114, FI-02540 Kylmälä, Finland}\par
{\footnotesize Max-Planck-Institut für Radioastronomie, Auf dem Hügel 69, D-53121 Bonn, Germany}\par
\vspace{0.55em}
\noindent\textbf{Karl-Friedrich Schuster} \hfill {\scriptsize ORCID: \texttt{0000-0003-2890-9454}}\par
{\footnotesize Institut de Radioastronomie Millimétrique (IRAM), 300 rue de la Piscine, F-38406 Saint Martin d'Hères, France}\par
\vspace{0.55em}
\noindent\textbf{Zhiqiang Shen (\cntext{沈志强})} \hfill {\scriptsize ORCID: \texttt{0000-0003-3540-8746}}\par
{\footnotesize Shanghai Astronomical Observatory, Chinese Academy of Sciences, 80 Nandan Road, Shanghai 200030, People's Republic of China}\par
{\footnotesize Key Laboratory of Radio Astronomy and Technology, Chinese Academy of Sciences, A20 Datun Road, Chaoyang District, Beijing, 100101, People’s Republic of China}\par
\vspace{0.55em}
\noindent\textbf{Sasikumar Silpa} \hfill {\scriptsize ORCID: \texttt{0000-0003-0667-7074}}\par
{\footnotesize Astronomy Department, Universidad de Concepción, Casilla 160-C, Concepción, Chile}\par
\vspace{0.55em}
\noindent\textbf{Randall Smith} \hfill {\scriptsize ORCID: \texttt{0000-0003-4284-4167}}\par
{\footnotesize Center for Astrophysics $|$ Harvard \& Smithsonian, 60 Garden Street, Cambridge, MA 02138, USA}\par
\vspace{0.55em}
\noindent\textbf{Bong Won Sohn} \hfill {\scriptsize ORCID: \texttt{0000-0002-4148-8378}}\par
{\footnotesize Korea Astronomy and Space Science Institute, Daedeok-daero 776, Yuseong-gu, Daejeon 34055, Republic of Korea}\par
{\footnotesize Korea National University of Science and Technology, Gajeong-ro 217, Yuseong-gu, Daejeon 34113, Republic of Korea}\par
\vspace{0.55em}
\noindent\textbf{Jason SooHoo} \hfill {\scriptsize ORCID: \texttt{0000-0003-1938-0720}}\par
{\footnotesize Massachusetts Institute of Technology Haystack Observatory, 99 Millstone Road, Westford, MA 01886, USA}\par
\vspace{0.55em}
\noindent\textbf{Kamal Souccar} \hfill {\scriptsize ORCID: \texttt{0000-0001-7915-5272}}\par
{\footnotesize Department of Astronomy, University of Massachusetts, Amherst, MA 01003, USA}\par
\vspace{0.55em}
\noindent\textbf{Joshua S. Stanway} \hfill {\scriptsize ORCID: \texttt{0009-0003-7659-4642}}\par
{\footnotesize Jeremiah Horrocks Institute, University of Lancashire, Preston PR1 2HE, UK}\par
\vspace{0.55em}
\noindent\textbf{He Sun (\cntext{孙赫})} \hfill {\scriptsize ORCID: \texttt{0000-0003-1526-6787}}\par
{\footnotesize National Biomedical Imaging Center, Peking University, Beijing 100871, People’s Republic of China}\par
{\footnotesize College of Future Technology, Peking University, Beijing 100871, People’s Republic of China}\par
\vspace{0.55em}
\noindent\textbf{Alexandra J. Tetarenko} \hfill {\scriptsize ORCID: \texttt{0000-0003-3906-4354}}\par
{\footnotesize Department of Physics and Astronomy, University of Lethbridge, Lethbridge, Alberta T1K 3M4, Canada}\par
\vspace{0.55em}
\noindent\textbf{Paul Tiede} \hfill {\scriptsize ORCID: \texttt{0000-0003-3826-5648}}\par
{\footnotesize Center for Astrophysics $|$ Harvard \& Smithsonian, 60 Garden Street, Cambridge, MA 02138, USA}\par
{\footnotesize Black Hole Initiative at Harvard University, 20 Garden Street, Cambridge, MA 02138, USA}\par
\vspace{0.55em}
\noindent\textbf{Remo P. J. Tilanus} \hfill {\scriptsize ORCID: \texttt{0000-0002-6514-553X}}\par
{\footnotesize Steward Observatory and Department of Astronomy, University of Arizona, 933 N. Cherry Ave., Tucson, AZ 85721, USA}\par
\vspace{0.55em}
\noindent\textbf{Michael Titus} \hfill {\scriptsize ORCID: \texttt{0000-0001-9001-3275}}\par
{\footnotesize Massachusetts Institute of Technology Haystack Observatory, 99 Millstone Road, Westford, MA 01886, USA}\par
\vspace{0.55em}
\noindent\textbf{Kenji Toma} \hfill {\scriptsize ORCID: \texttt{0000-0002-7114-6010}}\par
{\footnotesize Frontier Research Institute for Interdisciplinary Sciences, Tohoku University, Sendai 980-8578, Japan}\par
{\footnotesize Astronomical Institute, Tohoku University, Sendai 980-8578, Japan}\par
\vspace{0.55em}
\noindent\textbf{Pablo Torne} \hfill {\scriptsize ORCID: \texttt{0000-0001-8700-6058}}\par
{\footnotesize Institut de Radioastronomie Millimétrique (IRAM), Avenida Divina Pastora 7, Local 20, E-18012, Granada, Spain}\par
{\footnotesize Max-Planck-Institut für Radioastronomie, Auf dem Hügel 69, D-53121 Bonn, Germany}\par
\vspace{0.55em}
\noindent\textbf{Teresa Toscano} \hfill {\scriptsize ORCID: \texttt{0000-0003-3658-7862}}\par
{\footnotesize Instituto de Astrofísica de Andalucía-CSIC, Glorieta de la Astronomía s/n, E-18008 Granada, Spain}\par
{\footnotesize Max-Planck-Institut für Radioastronomie, Auf dem Hügel 69, D-53121 Bonn, Germany}\par
\vspace{0.55em}
\noindent\textbf{Efthalia Traianou} \hfill {\scriptsize ORCID: \texttt{0000-0002-1209-6500}}\par
{\footnotesize Instituto de Astrofísica de Andalucía-CSIC, Glorieta de la Astronomía s/n, E-18008 Granada, Spain}\par
{\footnotesize Max-Planck-Institut für Radioastronomie, Auf dem Hügel 69, D-53121 Bonn, Germany}\par
\vspace{0.55em}
\noindent\textbf{Sascha Trippe} \hfill {\scriptsize ORCID: \texttt{0000-0003-0465-1559}}\par
{\footnotesize Department of Physics and Astronomy, Seoul National University, Gwanak-gu, Seoul 08826, Republic of Korea}\par
{\footnotesize SNU Astronomy Research Center, Seoul National University, Gwanak-gu, Seoul 08826, Republic of Korea}\par
\vspace{0.55em}
\noindent\textbf{Matthew Turk} \hfill {\scriptsize ORCID: \texttt{0000-0002-5294-0198}}\par
{\footnotesize Department of Astronomy, University of Illinois at Urbana-Champaign, 1002 West Green Street, Urbana, IL 61801, USA}\par
\vspace{0.55em}
\noindent\textbf{Akhil Uniyal} \hfill {\scriptsize ORCID: \texttt{0000-0001-8213-646X}}\par
{\footnotesize Tsung-Dao Lee Institute, Shanghai Jiao Tong University, Shengrong Road 520, Shanghai, 201210, People’s Republic of China}\par
\vspace{0.55em}
\noindent\textbf{Ilse van Bemmel} \hfill {\scriptsize ORCID: \texttt{0000-0001-5473-2950}}\par
{\footnotesize ASTRON, Oude Hoogeveensedijk 4, 7991 PD Dwingeloo, The Netherlands}\par
\vspace{0.55em}
\noindent\textbf{Bram van den Berg} \hfill {\scriptsize ORCID: \texttt{0009-0000-9340-4204}}\par
{\footnotesize Department of Astrophysics, Institute for Mathematics, Astrophysics and Particle Physics (IMAPP), Radboud University, P.O. Box 9010, 6500 GL Nijmegen, The Netherlands}\par
\vspace{0.55em}
\noindent\textbf{Huib Jan van Langevelde} \hfill {\scriptsize ORCID: \texttt{0000-0002-0230-5946}}\par
{\footnotesize Joint Institute for VLBI ERIC (JIVE), Oude Hoogeveensedijk 4, 7991 PD Dwingeloo, The Netherlands}\par
{\footnotesize Leiden Observatory, Leiden University, Postbus 2300, 9513 RA Leiden, The Netherlands}\par
\vspace{0.55em}
\noindent\textbf{Daniel R. van Rossum} \hfill {\scriptsize ORCID: \texttt{0000-0001-7772-6131}}\par
{\footnotesize Department of Astrophysics, Institute for Mathematics, Astrophysics and Particle Physics (IMAPP), Radboud University, P.O. Box 9010, 6500 GL Nijmegen, The Netherlands}\par
\vspace{0.55em}
\noindent\textbf{Sebastiano D. von Fellenberg} \hfill {\scriptsize ORCID: \texttt{0000-0002-9156-2249}}\par
{\footnotesize Canadian Institute for Theoretical Astrophysics, University of Toronto, 60 St. George Street, Toronto, ON M5S 3H8, Canada}\par
{\footnotesize Max-Planck-Institut für Radioastronomie, Auf dem Hügel 69, D-53121 Bonn, Germany}\par
\vspace{0.55em}
\noindent\textbf{Jesse Vos} \hfill {\scriptsize ORCID: \texttt{0000-0003-3349-7394}}\par
{\footnotesize Centre for Mathematical Plasma Astrophysics, Department of Mathematics, KU Leuven, Celestijnenlaan 200B, B-3001 Leuven, Belgium}\par
\vspace{0.55em}
\noindent\textbf{Jan Wagner} \hfill {\scriptsize ORCID: \texttt{0000-0003-1105-6109}}\par
{\footnotesize Max-Planck-Institut für Radioastronomie, Auf dem Hügel 69, D-53121 Bonn, Germany}\par
\vspace{0.55em}
\noindent\textbf{Zhiren Wang} \hfill {\scriptsize ORCID: \texttt{0009-0004-9417-2213}}\par
{\footnotesize Perimeter Institute for Theoretical Physics, 31 Caroline Street North, Waterloo, ON N2L 2Y5, Canada}\par
{\footnotesize Department of Physics and Astronomy, University of Waterloo, 200 University Avenue West, Waterloo, ON N2L 3G1, Canada}\par
{\footnotesize Waterloo Centre for Astrophysics, University of Waterloo, Waterloo, ON N2L 3G1, Canada}\par
\vspace{0.55em}
\noindent\textbf{Derek Ward-Thompson} \hfill {\scriptsize ORCID: \texttt{0000-0003-1140-2761}}\par
{\footnotesize Jeremiah Horrocks Institute, University of Lancashire, Preston PR1 2HE, UK}\par
\vspace{0.55em}
\noindent\textbf{John Wardle} \hfill {\scriptsize ORCID: \texttt{0000-0002-8960-2942}}\par
{\footnotesize Physics Department, Brandeis University, 415 South Street, Waltham, MA 02453, USA}\par
\vspace{0.55em}
\noindent\textbf{Jasmin E. Washington} \hfill {\scriptsize ORCID: \texttt{0000-0002-7046-0470}}\par
{\footnotesize Steward Observatory and Department of Astronomy, University of Arizona, 933 N. Cherry Ave., Tucson, AZ 85721, USA}\par
\vspace{0.55em}
\noindent\textbf{Jonathan Weintroub} \hfill {\scriptsize ORCID: \texttt{0000-0002-4603-5204}}\par
{\footnotesize Center for Astrophysics $|$ Harvard \& Smithsonian, 60 Garden Street, Cambridge, MA 02138, USA}\par
{\footnotesize Black Hole Initiative at Harvard University, 20 Garden Street, Cambridge, MA 02138, USA}\par
\vspace{0.55em}
\noindent\textbf{Maciek Wielgus} \hfill {\scriptsize ORCID: \texttt{0000-0002-8635-4242}}\par
{\footnotesize Instituto de Astrofísica de Andalucía-CSIC, Glorieta de la Astronomía s/n, E-18008 Granada, Spain}\par
\vspace{0.55em}
\noindent\textbf{Kaj Wiik} \hfill {\scriptsize ORCID: \texttt{0000-0002-0862-3398}}\par
{\footnotesize Tuorla Observatory, Department of Physics and Astronomy, University of Turku, FI-20014 Turun Yliopisto, Finland}\par
{\footnotesize Finnish Centre for Astronomy with ESO, University of Turku, FI-20014 Turun Yliopisto, Finland}\par
{\footnotesize Aalto University Metsähovi Radio Observatory, Metsähovintie 114, FI-02540 Kylmälä, Finland}\par
\vspace{0.55em}
\noindent\textbf{Michael F. Wondrak} \hfill {\scriptsize ORCID: \texttt{0000-0002-6894-1072}}\par
{\footnotesize Anton Pannekoek Institute for Astronomy, University of Amsterdam, Science Park 904, 1098 XH, Amsterdam, The Netherlands}\par
{\footnotesize Gravitation and Astroparticle Physics Amsterdam (GRAPPA) Institute, University of Amsterdam, Science Park 904, 1098 XH Amsterdam, The Netherlands}\par
{\footnotesize Department of Astrophysics, Institute for Mathematics, Astrophysics and Particle Physics (IMAPP), Radboud University, P.O. Box 9010, 6500 GL Nijmegen, The Netherlands}\par
{\footnotesize Excellence Fellow at Radboud University, Nijmegen, The Netherlands}\par
\vspace{0.55em}
\noindent\textbf{George N. Wong} \hfill {\scriptsize ORCID: \texttt{0000-0001-6952-2147}}\par
{\footnotesize School of Natural Sciences, Institute for Advanced Study, 1 Einstein Drive, Princeton, NJ 08540, USA}\par
{\footnotesize Princeton Gravity Initiative, Jadwin Hall, Princeton University, Princeton, NJ 08544, USA}\par
\vspace{0.55em}
\noindent\textbf{Jompoj Wongphexhauxsorn} \hfill {\scriptsize ORCID: \texttt{0000-0002-7730-4956}}\par
{\footnotesize Julius-Maximilians-Universität Würzburg, Fakultät für Physik und Astronomie, Institut für Theoretische Physik und Astrophysik, Lehrstuhl für Astronomie, Emil-Fischer-Str. 31, D-97074 Würzburg, Germany}\par
{\footnotesize Max-Planck-Institut für Radioastronomie, Auf dem Hügel 69, D-53121 Bonn, Germany}\par
\vspace{0.55em}
\noindent\textbf{Qingwen Wu (\cntext{吴庆文})} \hfill {\scriptsize ORCID: \texttt{0000-0003-4773-4987}}\par
{\footnotesize School of Physics, Huazhong University of Science and Technology, Wuhan, Hubei, 430074, People's Republic of China}\par
\vspace{0.55em}
\noindent\textbf{Paul Yamaguchi} \hfill {\scriptsize ORCID: \texttt{0000-0002-6017-8199}}\par
{\footnotesize Center for Astrophysics $|$ Harvard \& Smithsonian, 60 Garden Street, Cambridge, MA 02138, USA}\par
\vspace{0.55em}
\noindent\textbf{Aristomenis Yfantis} \hfill {\scriptsize ORCID: \texttt{0000-0002-3244-7072}}\par
{\footnotesize Instituto de Astrofísica de Andalucía-CSIC, Glorieta de la Astronomía s/n, E-18008 Granada, Spain}\par
\vspace{0.55em}
\noindent\textbf{Doosoo Yoon} \hfill {\scriptsize ORCID: \texttt{0000-0001-8694-8166}}\par
{\footnotesize Anton Pannekoek Institute for Astronomy, University of Amsterdam, Science Park 904, 1098 XH, Amsterdam, The Netherlands}\par
\vspace{0.55em}
\noindent\textbf{André Young} \hfill {\scriptsize ORCID: \texttt{0000-0003-0000-2682}}\par
{\footnotesize Department of Astrophysics, Institute for Mathematics, Astrophysics and Particle Physics (IMAPP), Radboud University, P.O. Box 9010, 6500 GL Nijmegen, The Netherlands}\par
\vspace{0.55em}
\noindent\textbf{Ziri Younsi} \hfill {\scriptsize ORCID: \texttt{0000-0001-9283-1191}}\par
{\footnotesize Mullard Space Science Laboratory, University College London, Holmbury St. Mary, Dorking, Surrey, RH5 6NT, UK}\par
{\footnotesize Institut für Theoretische Physik, Goethe-Universität Frankfurt, Max-von-Laue-Straße 1, D-60438 Frankfurt am Main, Germany}\par
\vspace{0.55em}
\noindent\textbf{Wei Yu (\cntext{于威})} \hfill {\scriptsize ORCID: \texttt{0000-0002-5168-6052}}\par
{\footnotesize Institute for Astronomy, University of Hawai’i, 640 North Aohoku Place, Hilo, HI 96720, USA}\par
\vspace{0.55em}
\noindent\textbf{Feng Yuan (\cntext{袁峰})} \hfill {\scriptsize ORCID: \texttt{0000-0003-3564-6437}}\par
{\footnotesize Center for Astronomy and Astrophysics and Department of Physics, Fudan University, Shanghai 200438, People's Republic of China}\par
\vspace{0.55em}
\noindent\textbf{Ye-Fei Yuan (\cntext{袁业飞})} \hfill {\scriptsize ORCID: \texttt{0000-0002-7330-4756}}\par
{\footnotesize Astronomy Department, University of Science and Technology of China, Hefei 230026, People's Republic of China}\par
\vspace{0.55em}
\noindent\textbf{Ai-Ling Zeng (\cntext{曾艾玲})} \hfill {\scriptsize ORCID: \texttt{0009-0000-9427-4608}}\par
{\footnotesize Instituto de Astrofísica de Andalucía-CSIC, Glorieta de la Astronomía s/n, E-18008 Granada, Spain}\par
\vspace{0.55em}
\noindent\textbf{J. Anton Zensus} \hfill {\scriptsize ORCID: \texttt{0000-0001-7470-3321}}\par
{\footnotesize Max-Planck-Institut für Radioastronomie, Auf dem Hügel 69, D-53121 Bonn, Germany}\par
\vspace{0.55em}
\noindent\textbf{Shuo Zhang} \hfill {\scriptsize ORCID: \texttt{0000-0002-2967-790X}}\par
{\footnotesize Department of Physics and Astronomy, Michigan State University, 567 Wilson Rd, East Lansing, MI 48824, USA}\par
\vspace{0.55em}
\noindent\textbf{Brandon Zhao} \hfill {\scriptsize ORCID: \texttt{0009-0005-3991-9879}}\par
{\footnotesize California Institute of Technology, 1200 East California Boulevard, Pasadena, CA 91125, USA}\par
\vspace{0.55em}
\noindent\textbf{Guang-Yao Zhao} \hfill {\scriptsize ORCID: \texttt{0000-0002-4417-1659}}\par
{\footnotesize Max-Planck-Institut für Radioastronomie, Auf dem Hügel 69, D-53121 Bonn, Germany}\par
{\footnotesize Instituto de Astrofísica de Andalucía-CSIC, Glorieta de la Astronomía s/n, E-18008 Granada, Spain}\par
\vspace{0.55em}
\endgroup
\endgroup

\end{document}